\documentclass[twocolumn]{aastex63}
\usepackage{amsmath}
\usepackage{mathrsfs}
\submitjournal{ApJ}

\shorttitle{Populating the Black Hole Mass Gaps}
\shortauthors{Samsing \& Hotokezaka}

\begin{document}

\title{Populating the Black Hole Mass Gaps In Stellar Clusters:\\
General Relations and Upper Limits}

\author[0000-0003-0607-8741]{Johan Samsing}
\affiliation{Niels Bohr International Academy, The Niels Bohr Institute, \\
Blegdamsvej 17, 2100 Copenhagen, Denmark.}

\author{Kenta Hotokezaka}
\affiliation{Department of Astrophysical Sciences, Princeton University, \\
Peyton Hall, 4 Ivy Lane, Princeton, NJ 08544, USA.}
\affiliation{Research Center for the Early Universe, Graduate School of Science, \\
University of Tokyo, Bunkyo-ku, Tokyo 113-0033, Japan.}

\correspondingauthor{Johan Samsing}
\email{jsamsing@gmail.com}

\begin{abstract}

Theory and observations suggest that single-star evolution is not able to produce
black holes (BHs) with masses in the range $3-5M_{\odot}$ and above $\sim 45M_{\odot}$, referred to as the lower mass gap (LMG)
and the upper mas gap (UMG), respectively. However, it is possible to form BHs in these gaps through merger of compact objects in dense clusters, e.g. the LMG and the UMG can be populated through binary neutron star- and BBH mergers, respectively.
This implies that if binary mergers are observed in gravitational waves (GWs) with at least one mass gap object, then either
clusters are effective in assembling binary mergers, or our single-star models have to be revised.
Understanding how effective clusters are at populating both mass gaps have therefore major implications for both stellar- and GW astrophysics.
In this paper we present a systematic study on how efficient stellar clusters are at populating both mass gaps through in-cluster GW mergers.
For this, we derive a set of closed form relations for describing the evolution of compact object binaries undergoing dynamical interactions
and GW merger inside their cluster. By considering both static and time evolving populations, we find in particular that globular clusters are clearly inefficient at populating the
LMG in contrast to the UMG. We further describe how these results relate to the characteristic mass, time, and length scales associated with the problem.

\end{abstract}


\section{Introduction}\label{sec:Introduction}

Several gravitational wave (GW) sources have now been observed by the LIGO and Virgo GW observatories,
including both binary black holes (BBHs) \citep{2016PhRvL.116f1102A, 2016PhRvL.116x1103A, 2016PhRvX...6d1015A,
2017PhRvL.118v1101A, 2017PhRvL.119n1101A, 2019arXiv190210331Z, 2019arXiv190407214V}, and binary neutron stars (BNSs) \citep{2017PhRvL.119p1101A}.
Their astrophysical origin is still
unknown, but several formation channels have been suggested. Some of the recently proposed include:
field binaries \citep{2012ApJ...759...52D, 2013ApJ...779...72D, 2015ApJ...806..263D, Kinugawa2014MNRAS,2016ApJ...819..108B,
2016Natur.534..512B, 2017ApJ...836...39S, 2017ApJ...845..173M, 2018ApJ...863....7R, 2018ApJ...862L...3S},
dense stellar clusters \citep{2000ApJ...528L..17P,
2010MNRAS.402..371B, 2013MNRAS.435.1358T, 2014MNRAS.440.2714B,
2015PhRvL.115e1101R, 2016PhRvD..93h4029R, 2016ApJ...824L...8R,
2016ApJ...824L...8R, 2017MNRAS.464L..36A, 2017MNRAS.469.4665P, 2018PhRvD..97j3014S, 2018MNRAS.tmp.2223S, 2019arXiv190711231S},
active galactic nuclei (AGN) discs \citep{2017ApJ...835..165B,  2017MNRAS.464..946S, 2017arXiv170207818M, 2019arXiv191208218T},
galactic nuclei (GN) \citep{2009MNRAS.395.2127O, 2015MNRAS.448..754H,
2016ApJ...828...77V, 2016ApJ...831..187A, 2016MNRAS.460.3494S, 2017arXiv170609896H, 2018ApJ...865....2H},
very massive stellar mergers \citep{Loeb:2016, Woosley:2016, Janiuk+2017, DOrazioLoeb:2017},
and single-single GW captures of primordial black holes \citep{2016PhRvL.116t1301B, 2016PhRvD..94h4013C,
2016PhRvL.117f1101S, 2016PhRvD..94h3504C}.

The question is, which of these proposed merger channels dominate the merger rate? Are several channels operating with a possible dependence on redshift? Or
are the majority of GW sources formed through a still unknown mechanism? Several studies show that one can
distinguish at least classes of channels apart, such as isolated binaries and dynamically induced mergers,
by considering the observed distribution of merger masses \citep{2017ApJ...846...82Z}, the relative spin orientation of the merging objects \citep{2016ApJ...832L...2R},
as well as the orbital eccentricity at some
reference GW frequency \citep{2006ApJ...640..156G, 2014ApJ...784...71S, 2017ApJ...840L..14S, 2018MNRAS.476.1548S, 2018ApJ...853..140S, 2018PhRvD..97j3014S, 2018ApJ...855..124S,
2018MNRAS.tmp.2223S, 2018PhRvD..98l3005R, 2019ApJ...871...91Z, 2019PhRvD.100d3010S, 2019arXiv190711231S}. Other `indirect' probes have also been suggested,
such as stellar tidal disruptions \citep[e.g.][]{2019PhRvD.100d3009S, 2019ApJ...877...56L, 2019ApJ...881...75K}.
In this picture, it is now largely believed that dynamically assembled mergers are likely to have mass rations near one \citep[e.g.][]{2018PhRvD..98l3005R}, random
relative spin orientations \citep[e.g.][]{2016ApJ...832L...2R}, and a non-negligible fraction of mergers with measurable eccentricity in
both LISA \citep{2018MNRAS.tmp.2223S, 2019PhRvD..99f3003K}, DECIGO/Tian-Qin \citep[e.g.][]{2017ApJ...842L...2C, 2019arXiv190711231S}, and LIGO \citep{2018PhRvD..97j3014S}.
This is in contrast to isolated binary mergers, that likely have correlated spins \citep[e.g.][]{2000ApJ...541..319K}, a bimodal distribution  for the effective spin parameter \citep{Zaldarriaga2018MNRAS,Hotokezaka2017ApJ,Piran2020ApJ},
larger mass ratios, and merge on orbits with eccentricities indistinguishable from $\approx 0$ near LISA and LIGO.
This picture is rather clean when comparing mergers forming in highly dynamical systems, such as
globular clusters (GCs) and GN, to completely isolated field binary mergers; however, it becomes less clean when considering e.g. the proposed sub-population of field binaries that undergo secular interactions
with nearby single or binary objects \citep[e.g.][]{2013ApJ...773..187N, 2016ARA&A..54..441N, 2016ComAC...3....6T, 2017ApJ...841...77A, 2017ApJ...836...39S, 2018ApJ...863...68L, 2018ApJ...863....7R, 2018ApJ...864..134R, 2018MNRAS.480L..58A, 2019MNRAS.483.4060L, 2019MNRAS.486.4443F, 2019MNRAS.486.4781F, 2019ApJ...883...23H, 2020ApJ...888L...3S}. In this case, secular exchanges of especially angular momentum, can drive the binary to merge with random spin orientations \citep[e.g.][]{2017ApJ...846L..11L}, and
notable eccentricity \citep[e.g.][]{2018ApJ...853...93R, 2019ApJ...881...41L, 2020MNRAS.493.3920F}, which makes it more challenging to disentangle cluster mergers from field binary mergers.
 
An additional outcome that is somewhat unique to dynamically environments is the formation of so-called hierarchical mergers \citep[e.g.][]{2016ApJ...824L..12O, 2017ApJ...840L..24F, 2017PhRvD..95l4046G, 2019PhRvL.123r1101Y, 2019MNRAS.486.5008A, 2019PhRvD.100d1301G, 2019MNRAS.482...30S, 2019PhRvD.100d3027R, 2020arXiv200504243G, 2020ApJ...888L...3S, 2020ApJ...890L..20G, 2020arXiv200500023K, 2020ApJ...893...35D, 2020arXiv200400650B}.
The picture is here that compact objects (COs) that merge inside their cluster through e.g. single-single GW captures \citep[e.g.][]{2019arXiv190711231S} or through
chaotic few-body interactions \citep[e.g.][]{2014ApJ...784...71S, 2019ApJ...871...91Z}, will form a new population of `second-generation' (2G) objects that are
characterized by having a higher mass than the original `first-generation' (1G) population, and a dimensional spin parameter around 0.7 \citep[e.g.][]{2007PhRvD..76f4034B}.
This 2G population can undergo further interactions leading to merger with other 1G or 2G objects, which then naturally will
lead to an observable modified BBH mass spectrum, and spin distribution. This process can in principle also lead to 3G-, 4G-, ..., $N$G-populations, which naturally gives rise to
unique observables. Looking for such hierarchical merger configurations has been proposed to be one way of probing the origin of GW mergers in very dense
systems, such as GCs \citep{2019PhRvD.100d3027R}, GN \citep{2016ApJ...831..187A}, and AGN disks \citep{2019PhRvL.123r1101Y}. However,
fine-tuned few-body configurations in the binary field population can in principle also create hierarchical mergers \citep[e.g.][]{2020ApJ...888L...3S},
but in this case its highly unlikely to go beyond 2G. In any case, an observation of a hierarchical merger would strongly indicate that at least some GW sources are assembled as a result of
few-body interactions.

Another interesting consequence of the hierarchical merger scenario is the possibility of populating the so-called lower mass gap (LMG) and upper mass gap (UMG), where
the LMG is $\sim 3M_{\odot} - 5M_{\odot}$ \citep{Bailyn1998ApJ,Ozel2010ApJ,farr2011ApJ} and the UMG is marked by a lower limit of $\sim 45M_{\odot}$ \citep{Woosley2017ApJ,Leung2019ApJ,Farmer2019ApJ}. For example, the LMG might be populated through BNS collisions,
where the UMG can be populated by BBH mergers. This makes it possible for dense clusters to produce GW sources with objects in either the LMG or the UMG. If `Nature' is not able to form BHs through single star evolution in these mass gaps, then an observation of GW sources
with a mass-gap object will give us insight into the fraction of mergers assembled in clusters, or at least dynamically.
These mass-gaps not only play a key role in stellar-astrophysics, but introduce also a
characteristic mass scale that can be used to e.g. constrain the cosmological parameters \citep{2019ApJ...883L..42F}.

Several recent studies have discussed the possibility of populating the mass-gap in clusters \citep[e.g.][]{2019PhRvD.100d3027R, 2020ApJ...893...35D, 2020arXiv200400650B}.
Currently, numerical studies suggest that BNS mergers are not likely to form in
systems such as GCs \citep[e.g.][]{2020ApJ...888L..10Y}. On the other hand, recent observations of the orbital parameters of galactic BNSs
interestingly indicate that BNSs might actually form in clusters at rates several orders-of-magnitude higher than suggested by the numerical studies \citep[e.g.][]{2019ApJ...880L...8A},
which of course poses some interesting tension. Regarding BBHs, several studies have found that if the initial BH spins are low, then up to $\sim 10\%$ of BBH mergers from GCs could be in the form of 1G-2G binaries, with a sub-fraction of these being in
the UMG \citep{2019PhRvD.100d3027R}. To find the observable contribution from such hierarchical mergers
in upcoming and future GW data several numerical techniques and models are now under development \citep[e.g.][]{2020ApJ...893...35D}; however, common for the majority of
these models is that they are not linked to any real physical model, they are instead just generic functional forms with a few fitting parameters. This kind of model independent
approach might be useful to condense a huge stream of data into just a few fitting parameters, but gives a-priori no astrophysical insight into what
systems that are likely and able to undergo hierarchical mergers and populating the mass-gaps.

In this paper we derive a set of fundamental relations describing how effective a dense cluster can grow a 2G-population from a series of in-cluster GW mergers of 1G-1G binaries,
as a function of characteristic mass, length, and time scales of the 1G objects and their cluster. The core of our calculations are based on the post-Newtonian (PN)
binary-single hardening model presented in \cite{2018PhRvD..97j3014S, 2018MNRAS.tmp.2223S}, where binaries are able to
merge in-between or during their hard binary-single interactions.
We use our derived expressions to make general statements about what clusters that are able to populate the LMG through BNS mergers, and the UMG through BBH mergers.
Our model is fully analytical and our results are given in closed form expressions, and as a result, we are therefore only able to describe idealized clusters with constant
density and velocity dispersion (for an extension of our model see e.g. \citealt{2020MNRAS.492.2936A}); however, our work serves as an important first
step in connecting physical parameters with more general statements related to hierarchical mergers (see also recent work by \citealt{2020arXiv200400650B}).

The paper is organized as follows. In Sec. \ref{sec:Binary Burning and In-Cluster Mergers} we
introduce our dynamical cluster- and 3-body interaction model, and use it to derive results on how efficient a simple non-evolving binary and single cluster population is at producing
in-cluster GW mergers. In Sec. \ref{sec:Populating the Black Hole Mass Gaps: Time Evolving Cluster Model} we extend our model to include a
time dependent distribution of both singles and binaries, from which we derive a closed form solution to the upper limit on the number of $2$G-objects
relative to $1$G-objects a given cluster can reach in a Hubble time. We further discuss these results in relation to populating the LMG and the UMG. We conclude our study in Sec. \ref{sec:Conclusions}.

\section{Formation of In-Cluster Mergers}
\label{sec:Binary Burning and In-Cluster Mergers}

We consider a cluster with a population of COs (NSs or BHs), each with a mass $m$.
These COs interact, and can through different dynamical pathways merge through the emission
of GWs either inside or outside of their cluster \citep[e.g.][]{2018MNRAS.tmp.2223S, 2018PhRvD..98l3005R}.
The COs that merge inside the cluster give rise to a growing in-cluster population of BHs with a mass $ \approx 2m$, given that the
kick velocity associated with asymmetric GW emission at merger is smaller than the cluster escape velocity \citep[e.g.][]{2019PhRvD.100d1301G}.
In this work we refer to the initial population of COs by `1' or 1. generation (1G) objects, and the population of
BHs that is formed through the collision of 1G-1G binaries by `2', or 2. generation (2G).
As described in the Introduction, the 2G population is able to populate both the lower ($3-5 M_{\odot}$) and upper ( $> 45 M_{\odot}$)
BH mass gaps that are believed to be associated with the initial 1G population. For example, it might be possible to populate
the $3-5 M_{\odot}$ BH mass-gab through the collision of BNSs.

Below we start by deriving and present a set of basic relations for describing the growth of 2G populations
through in-cluster 1G-1G GW mergers.
Throughout the paper we mainly illustrate results for our two fiducial mass cases; $m=1.4M_{\odot}$ and $m = 30M_{\odot}$,
which are in the relevant range for populating the LMG and the UMG, respectively.

\subsection{Cluster Model and 3-Body Dynamics}
\label{sec:Cluster Model}

In this work we study a model described by a cluster consisting of COs all with the same mass $m$.
The cluster itself is assumed to have a constant number density of singles $n$,
velocity dispersion $v_d$, and escape velocity $v_e = f_{ed} \times v_d$. Besides this single population,
the cluster also harbors a population of CO binaries, that at early times consist of 1G-1G pairs,
but at later times, through dynamical exchange interactions, can evolve to have pairs also including 2G objects.
The binaries play a very important role, as these provide the main pathway for producing 2G-objects as a result of 
binary-single interactions. In-cluster GW mergers can also form in other
ways, such as through single-single GW captures \citep{2019arXiv190711231S}, secular Kozai triples \citep{2016ApJ...816...65A}, and binary-binary interactions \citep{2019ApJ...871...91Z};
however, these pathways are generally subdominant compared to the binary-single channel. Our main discussions will therefore mostly
involve mergers from the interacting binary-single population. In the sections below we continue by describing the basics of our cluster model.

\subsubsection{Binary Hardening and Outcomes}
\label{sec:Binary Hardening and Outcomes}

\begin{figure}
\centering
\includegraphics[width=\columnwidth]{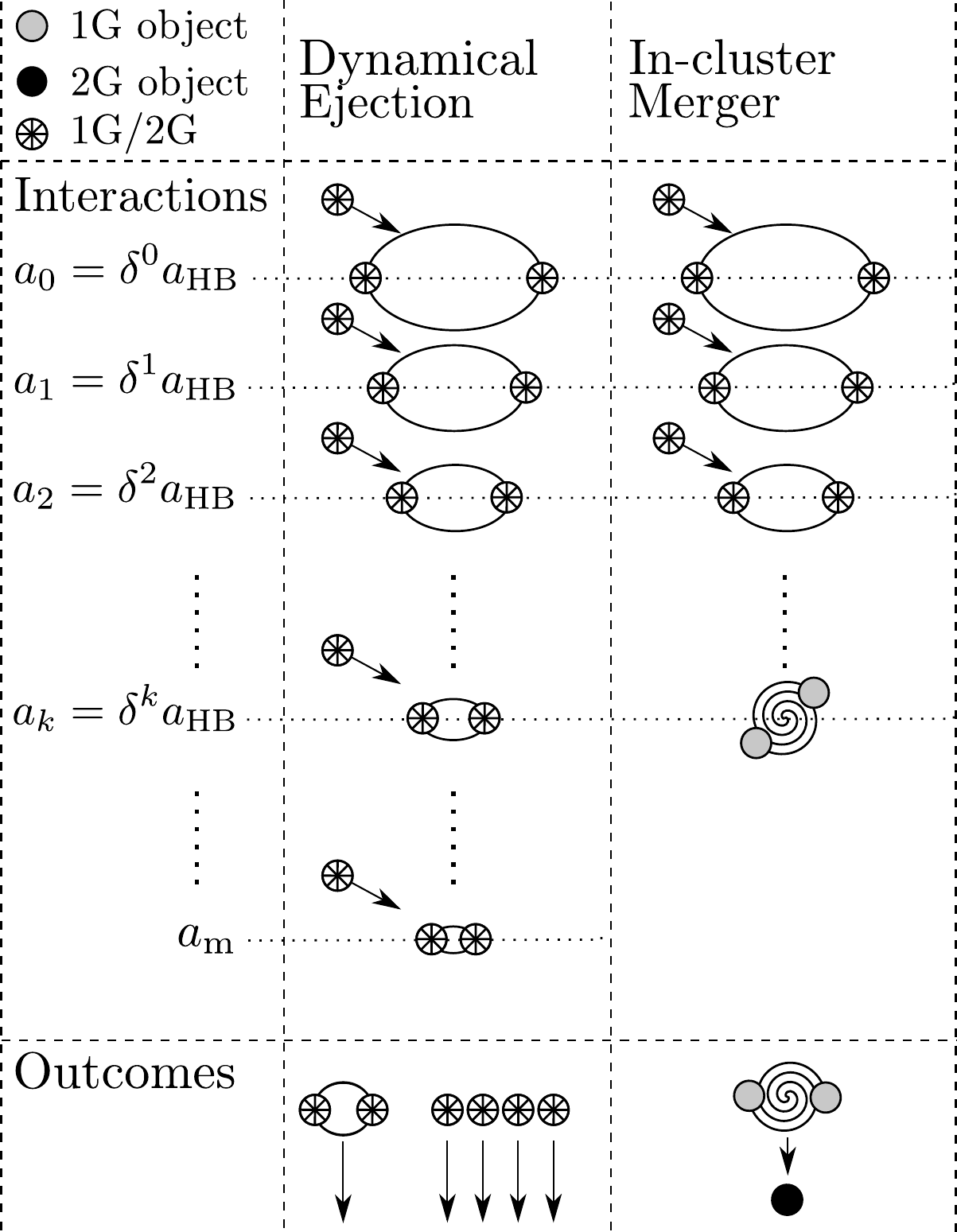}
\caption{Illustration of our $\delta$-model described in Sec. \ref{sec:Cluster Model}.
In this model we assume that all binaries dynamically form at the HB-limit inside the cluster, after which they
undergo scatterings with the surrounding singles. Each scattering leads to a fixed decrease in the SMA
from $a$ to $\delta a$, as shown in the left column ({\it Interactions}).
This series of hardening interactions terminate at a characteristic SMA $a_m$, that 
either is where the binary mergers inside the cluster, is being ejected, or when the time passes $t_H$, as further described in Sec. \ref{sec:Binary Hardening and Outcomes}.
The middle and the right columns show the two scenarios where the binary is either ejected ({\it Dynamical Ejection}), or
merges inside the cluster ({\it In-cluster Merger}), respectively. The outcome from each of these scenarios is shown in the bottom panel ({\it Outcomes}).
As seen, the outcome from a dynamical ejection is $1$ binary and $\sim 4$ singles, where for 1G-1G in-cluster mergers the outcome is per defintion a 2G-object.
When following the in-cluster population of 1G and 2G objects over time, the {\it Dynamical Ejection} outcome always acts as a `sink term', where
the 1G-1G {\it In-cluster Merger} outcome is the `source term' for the 2G population, as further described in Sec. \ref{sec:Evolution Equations}. 
Note that the {\it grey} and {\it black} circles refer to 1G- and 2G-objects, respectively, where the {\it diamond symbol} denotes either of these two objects.}
\label{fig:ill_int}
\end{figure}

We assume that a given CO binary inside the cluster forms (dynamically) with a semi-major (SMA), $a$, equal to the hard-binary (HB) limit
value \citep[e.g.][]{Heggie:1975uy, 1976A&A....53..259A, Hut:1983js},
\begin{equation}
a_{HB} = \frac{3G}{2} \times \frac{m}{v_d^2},
\end{equation}
which is where the binary binding energy ($Gm^2/(2a)$) equals the kinetic energy of the surrounding singles w.r.t. the binary ($mv_{d}^{2}/3$).
After this, the binary undergoes scatterings with the surrounding singles,
each of which leads to a decrease in the SMA of the binary from $a$ to $\delta a$. This
corresponds to a change in $a$ of $-a(1-\delta) = -a\Delta$, where we have introduced
$\Delta \equiv 1-\delta$ to shorten notations.
In reality, the change per interaction in the
binary binding energy $E_{b}$ follows approximately a power-law distribution $P(E_{b}) \propto E_b^{-\gamma}$
with $\gamma \sim 9/2$, depending on exactly how a strong binary-single interaction is defined \citep[e.g.][]{Heggie:1975uy, 2019Natur.576..406S}.
In this work we do not use the full distribution, instead we assume that each interaction leads to
a fixed fractional decrease $\delta$ in the SMA, that is equal to the average value found from the distribution $P(E_{b}) \propto E_b^{-\gamma}$.
Using that $E_b = -Gm^2/(2a)$, and that $\delta \equiv \langle a \rangle/a_0$, where $a_0$ is the initial SMA and $\langle a \rangle$ is the
average value of the resulting SMA, then $\delta$ is given by (see also \citealt{2018PhRvD..97j3014S}),
\begin{align}
	\delta &= (\gamma - 1) \int_{0}^{1} {\delta}^{(\gamma - 1)} d\delta = 1-{\gamma}^{-1}, \nonumber\\
	 	 &= {7}/{9},\ \ (\gamma = 9/2).
    \label{eq:delta}
\end{align}
The binary keeps undergoing these so-called `hardening' interactions with the surrounding single population, until its SMA reaches one of the
following three characteristic values:
The first, denoted by $a_{ej}$, is the maximum SMA value for which the binary will get ejected out of the cluster if it undergoes a
binary-single interaction. Note that this is a fixed value in our simple `$\delta$-model'.
The second, denoted by $a_{GW}$, is the SMA for which the total integrated probability for the binary to merge at any given state from
$a_{HB}$ to $a_{GW}$ equals one. The merging binary will of course have a decreasing SMA as it inspirals, but will
during this time not interact with other objects.
The third, denoted by $a_{tH}$, is the value it takes a Hubble time to reach through
binary-single interactions alone from the initial value $a_{HB}$.
The hierarchy of these three characteristic scales is set by $(v_d,n,m)$, and plays a key role in how to grow a 2G BH population
inside the cluster through in-cluster mergers (see also \citealt{2016ApJ...831..187A, 2020MNRAS.492.2936A, 2020arXiv200400650B}). For example,
if $a_{ej} > a_{GW}$ then most binaries will get ejected and merge outside of the cluster, compared to if $a_{GW} > a_{ej}$ in which case
all binaries will merge inside. If on the other hand $a_{tH} > \{a_{ej}, a_{GW}\}$ then the system will not be able to
conclude even a single interaction sequence, and an effective accumulation of 2G mergers is therefore near impossible.
As a result, the `relevant' value for a given system is
\begin{equation}
a_{m} = max(\{a_{ej}, a_{GW}, a_{tH}\}),
\label{eq:amin}
\end{equation}
where the sub-script `$m$' here refers to `minimum', as this is the smallest value the SMA of the interacting binary can take.
This $\delta$-model is further illustrated and described in Fig. \ref{fig:ill_int}.

Throughout the paper we refer to the process in which the system brings a binary from $a_{HB}$ to $a_m$ through binary-single
interactions alone as one `Interaction Cycle' (IC). After a binary has completed its IC, then it generally happens that a new binary dynamically
forms with a SMA $\sim a_{HB}$, after which the process repeats. This cycle of binary formation and hardening interactions is also often referred
to as `binary burning' \citep[e.g.][]{2020IAUS..351..357K}. We continue below by deriving $a_{ej}$, $a_{GW}$, and $a_{tH}$.
We also refer the reader to \citep{2016ApJ...831..187A, 2020MNRAS.492.2936A, 2020arXiv200400650B} for complementary discussions on this.

\subsubsection{Derivation of Outcome Conditions}
\label{sec:Overview...1}

For calculating the SMA at which the binary is ejected, $a_{ej}$, we first use that the
energy released in one interaction between a single and a binary with SMA $a$ is given by ${E_{bs}} = (\Delta/\delta) \times E_{b}(a)$, where
$E_{b}(a)$ is the internal energy of the binary before interaction \citep[e.g.][]{2018PhRvD..97j3014S}. The energy ${E_{bs}}$ is `released' in the three-body center-of-mass (COM),
which in the Newtonian limit is conserved from before to after the interaction. From momentum conservation it then follows that the binary receives
a velocity kick, defined at infinity in the COM, of $v_{b}^2 = {E_{bs}}/(3m) = (1/6)(\Delta/\delta)Gm/a$. When $a$ is
such that $v_{b} > v_{e}$ then the binary escapes the cluster. By now defining $a_{ej} \equiv a(v_{b} = v_{e})$
it then follows that,
\begin{equation}
a_{ej} \approx \frac{G\Delta}{6{\delta}{f_{ed}^2}} \times \frac{m}{v_{d}^2}.
\label{eq:aej}
\end{equation}
Note here that $a_{HB}/a_{ej} = 9f_{ed}^2{\delta}/{\Delta} = (63/2)f_{ed}^2$, where we have set $\delta = 7/9$ in the last
equality. A single binary therefore has to decrease its SMA by 1-2 orders of magnitude through binary-single scatterings
before a possible ejection can take place. As will be discussed and used later, several of the single objects
interacting with the binary will also get ejected, as they likewise receive recoil kicks during the hardening process.
As a result, for every single binary ejected there will also be $N^{ej}_{s}$ single objects ejected.
This number $N^{ej}_{s}$ can be estimated by first comparing the SMA below which
single ejections are possible, $a^{s}_{ej} \approx {2G\Delta}/({3{\delta}{f_{ed}^2}}) \times {m}/{v_{d}^2}$,
where we have used $2v_{b} = v_{s}$, with the binary ejection SMA $a_{ej}$ from Eq. \eqref{eq:aej}.
As seen, $a^{s}_{ej}/a_{ej} = 4$. Now using that after $\Delta{n}$ binary-single interactions the binary SMA
decreases by a factor $\delta^{\Delta{n}}$, it then follows that $N^{ej}_{s} = \ln{(1/4)}/\ln{\delta} \approx 5$,
where we have used that one single object is ejected in each scattering for $a_{ej} \leq a \leq a^{s}_{ej}$.
The number $N^{ej}_{s}$ is therefore a constant that does
not depend on any properties of the system, as long all the interaction steps are `available'. In this paper we
use $N^{ej}_{s} = 4$, as this value is slightly closer to what is found in numerical simulations; however, the exact value does
not play a large role, the important point is that it takes a constant value.

For $a_{GW}$, we start by calculating the probability that a binary with SMA $a$ merges before its next binary-single
interaction, denoted here by $p_{2}(a)$. For this we assume that the eccentricity distribution of the binary follows that of a
thermal distribution, $P(e)=2e$ \citep[e.g.][]{Heggie:1975uy}. In addition, we use that the time in-between binary-single interactions, $t_{bs}(a)$, is the inverse of the
binary-single encounter rate, $t_{bs}(a) \approx (n\sigma_{bs}v_{d})^{-1}$, where $\sigma_{bs} \propto ma/v_d^{2}$
is the HB binary-single interaction cross section (see e.g. \citealt{2018ApJ...853..140S}).
Under these assumptions it directly follows that $p_{2}(a) = (t_{bs}(a)/t_{GW}(a))^{2/7}$,
where $t_{GW}(a) \propto a^{4}/m^{3}$ is the GW inspiral life time corresponding to $e = 0$ \citep[e.g.][]{2018PhRvD..97j3014S}.
This $p_{2}(a)$ is only the probability for merger during a single `interaction step' $k$, where we here introduce the notation $a_{k} = a_{HB}{\delta}^{k}$. The total probability for a binary to merge in-between its binary-single interactions from $a_{HB}$ to $a_m$, denoted by $P_2(a_m)$, is therefore
found by simply integrating from $k(a_{HB}) = 0$ to $k(a_m)$. Using that $da = -a{\Delta} dk$, the solution is found to
be $P_2(a_m) \approx p_{2}(a_m) \times ({7}/({10{\Delta}}))$ \citep[e.g.][]{2018PhRvD..97j3014S, 2019PhRvD.100d3009S}, which can be written out as,
\begin{equation}
P_{2}(a_m) \approx A_c^{2/7} \times \frac{m^{4/7}{v_d}^{2/7}}{{n}^{2/7}{a_m}^{10/7}},
\label{eq:P2}
\end{equation}
where we have assumed that $p_2(a_m) \gg p_2(a_{HB})$ and defined the constant $A_c = ({7^{7/2}}{85G^2})/({(10\Delta)^{7/2}}{9{\pi}c^5})$.
If we now set $P_2 = 1$ then the corresponding $a_{GW} \equiv a(P_2 = 1)$ can now be
isolated and gives,
\begin{equation}
a_{GW} = A_c^{1/5} \times \frac{m^{2/5}{v_d}^{1/5}}{{n}^{1/5}}.
\label{eq:aGW}
\end{equation}
As seen, this limit is surprisingly insensitive to the cluster parameters $v_d$ and $n$ (see also \citealt{2016ApJ...831..187A}).

The last characteristic SMA we consider is $a_{tH}$, which is the value for which it takes the binary a Hubble time to
reach from $a = a_{HB}$ through binary-single interactions alone. For calculating this, we start with the time
it takes the binary to undergo one interaction at interaction step `$k$', which can be approximated as $t_{bs}(a_{k}) \approx (n\sigma_{bs}(a_k)v_{d})^{-1}$ (see the
above paragraph). The total time it takes to reach $a_m$, denoted by $\tau_m$, is found by integration $t_{bs}(a_{k})$
from $k(a_{HB}) = 0$ to $k(a_m)$. From this, one finds that $\tau_m \approx t_{bs}(a_m)/{\Delta}$, which also can be written as,
\begin{equation}
\tau_m \approx {\left( 6 \pi G {\Delta} \right)^{-1}} \times \frac{1}{a_m}\frac{v_d}{nm},
\label{eq:tauaf}
\end{equation}
where we have assumed that $a_{HB} \gg a_m$.
Setting this expression for $\tau_m$ equal to $t_H$, and isolating the corresponding $a_{tH} \equiv a_m(\tau_m = t_H)$, one now finds,
\begin{equation}
a_{tH} \approx \frac{{\left( 6 \pi G {\Delta} \right)^{-1}}}{t_H} \times \frac{v_d}{nm},
\label{eq:atH}
\end{equation}
which relates to $a_{HB}$ as $a_{HB}/a_{tH} \approx (t_{H}/t_{bs}(a_{HB})){\Delta}$.

\subsection{Results}

Having derived analytical expressions for the three characteristic scales
$a_{ej}$, $a_{GW}$, and $a_{tH}$ in Sec. \ref{sec:Overview...1} above, we are now in a position to start exploring what cluster systems that are likely to
grow a population of 2G objects. In the sections below we study this by considering a few general relations and
overview figures for a `static' cluster population. In Sec. \ref{sec:Populating the Black Hole Mass Gaps: Time Evolving Cluster Model}
we use these results to model `time evolving' populations.

\begin{figure}
\centering
\includegraphics[width=\columnwidth]{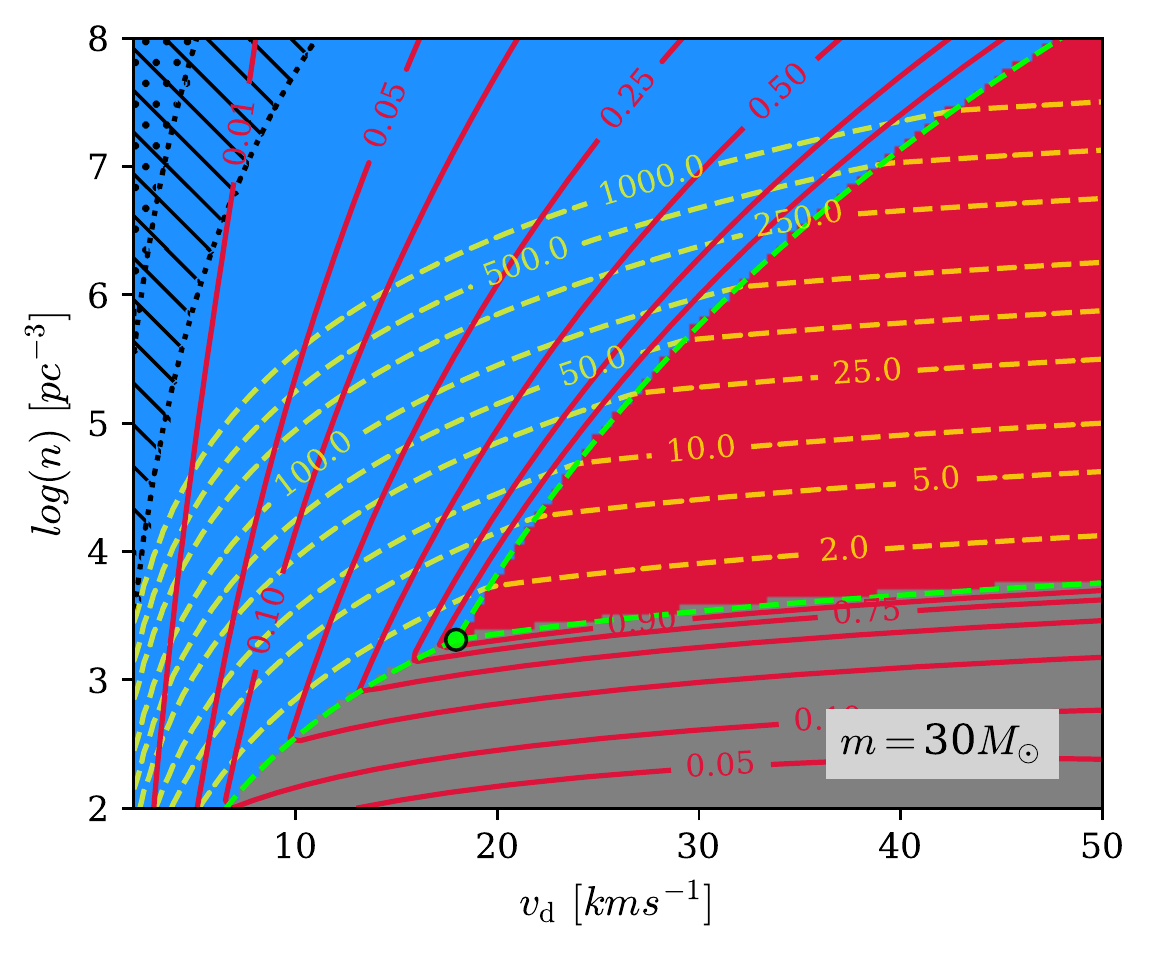}
\includegraphics[width=\columnwidth]{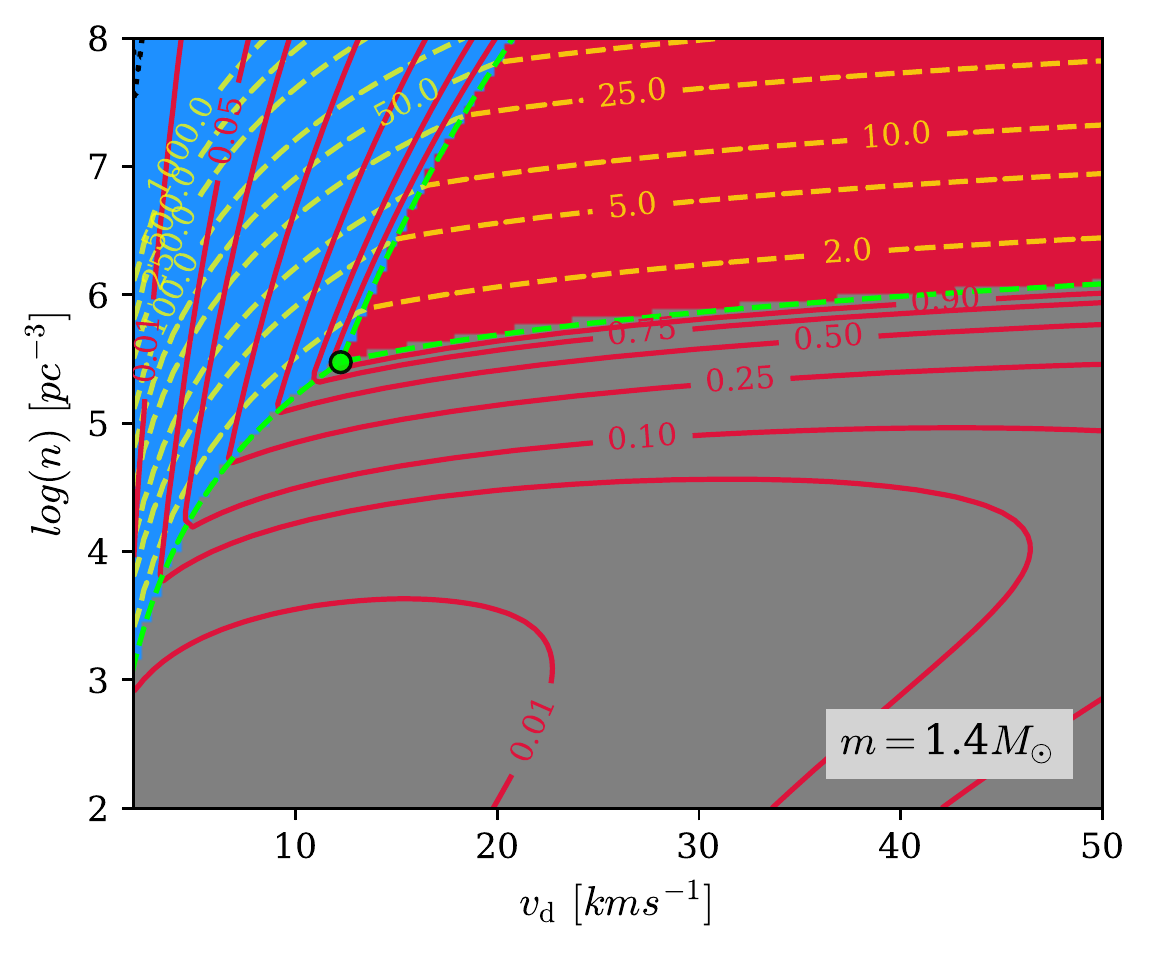}
\caption{Results for a binary interacting with singles in
a cluster described by a velocity dispersion $v_{d} = v_{e}/\sqrt{12}$, and number density $n$. The objects all have the same mass $m$,
where $m=30M_{\odot}$ (BH) and $m=1.4_{\odot}$ (NS) in the {\it top plot} and the {\it bottom plot}, respectively.
The binary evolution is modeled using our $\delta$-model described in Sec. \ref{sec:Cluster Model}.
In this model, the IC outcome of a given binary will fall into one of the following three general categories: {\it (1) Blue region}: The binary
can get ejected through a binary-single interaction ($a_m = a_{ej}$). {\it (2) Red region}: The binary will always merge inside the cluster
before ejection is possible ($a_m = a_{GW}$). {\it (3) Grey region}: The binary will not be able to finish a single IC
within a Hubble time ($a_m = a_{tH}$). The {\it green dashed lines} separating these outcome regions
are discussed in Sec. \ref{sec: Outcome Regions}.
The {\it red solid lines} show the total probability for a binary to merge inside the cluster
during one IC, $P_{M} = P_2 + P_3$, where the {\it orange dashed lines}
show the number of ICs a binary can undergo in Hubble time, $N_c(t_H)$.
The {\it hatched region} shows where $P_3>P_2$, where the {\it dotted region} shows where $\Gamma_{ss} > \Gamma_{23}$ for $N_b/N_s = 0.05$,
as further described in Sec. \ref{sec:Contribution from 3-body Mergers and S-S Captures}.}
\label{fig:alim_fig}
\end{figure}

\subsubsection{Outcome Regions}
\label{sec: Outcome Regions}

Fig. \ref{fig:alim_fig} shows $a_m = max(\{a_{ej}, a_{GW}, a_{tH}\})$ with colored regions ({\it blue}, {\it red}, {\it grey}) as a function
of cluster velocity dispersion $v_d$ and number density $n$ for $m = 30M_{\odot}$ (top) and $m = 1.4M_{\odot}$ (bottom).
The three regions are separated by {\it green dashed lines}, where the point at which all of the three regions meet, a point we refer to the `break point' (BP),
is highlighted with a {\it green circular dot}.
How the green dashed lines depend on the parameters $v_d, n, m$,
provide the key to understand what systems that are likely to produce a sizable population of 2G objects.
Below we study this in more detail. Our expressions are written out for $\delta = 7/9$ if nothing else is stated.

We start by the line separating the two regions $a_{ej}$ (blue) and $a_{tH}$ (grey) to the left of the BP. By now setting
$a_{ej} = a_{tH}$ and solving for the corresponding $n(ej,tH) \equiv n(a_{ej} = a_{tH})$ one finds,
\begin{equation}
n(ej,tH) = \left[\frac{63}{4\pi}\frac{f_{ed}^2}{G^2t_{H}}\right] \times \frac{v_{d}^3}{m^2}, 
\end{equation}
where we used Eq. \eqref{eq:aej} and \eqref{eq:atH}.
The next line is the one separating the regions $a_{ej}$ (blue) and $a_{GW}$ (red) to the right of the BP. Following the same procedure as
above, we first set $a_{ej} = a_{GW}$ and solve for the corresponding $n(ej,GW)$,
\begin{equation}
n(ej,GW) = \left[\frac{1}{B_c} \frac{f_{ed}^{10}}{G^3 c^5}\right] \times \frac{v_{d}^{11}}{m^3}, 
\end{equation}
where we have used Eq. \eqref{eq:aej} and \eqref{eq:aGW}, and introduced the constant $B_c = ({9 \pi}/({21^{5}}85))(20/63)^{7/2}$.
Finally, the line separating $a_{GW}$ (red) and $a_{tH}$ (grey) to the right of the BP is found from setting $a_{GW} = a_{tH}$,
from which we find,
\begin{equation}
n(GW,tH) = \left[\left(\frac{63}{4\pi}\right)^5 \frac{B_c c^5}{{G^7}{t_{H}^5}}\right]^{1/4} \times \frac{v_{d}}{m^{7/4}},
\end{equation}
where we have used Eq. (\ref{eq:aGW}) and (\ref{eq:atH}).
The associated coordinates of the BP, denoted by $(v_d(BP), n(BP))$, can now be
found from, e.g., setting $n(ej,tH) = n(ej,GW)$ from which follows,
\begin{align}
    v_d(BP) &= \left[\frac{63}{4\pi} \frac{B_c G c^5}{f_{ed}^8 t_H}\right]^{1/8} \times m^{1/8}, \nonumber\\
    n(BP)     &= \left[\left(\frac{63}{4\pi}\right)^{11} \frac{B_c^3 c^{15}}{G^{13}f_{ed}^8 t_H^{11}}\right]^{1/8} \times m^{-13/8}.
    \label{eq:BPc}
\end{align}
As seen here, the BP coordinates $v_d(BP),n(BP)$ are $\propto m^{1/8}, m^{-13/8}$, respectively. Therefore, the location of the BP along the
$v_d$-axis remains almost constant for reasonable changes in $m$,  in contrast to the location along the $n$-axis,
which can change by orders-of-magnitude. As a result, for 1G objects in the mass range $1 M_{\odot} \lesssim m \lesssim 50 M_{\odot}$
the BP will always be around $10 \sim 20\ kms^{-1}$, which is slightly higher than the dispersion velocity of a typical GC. Since no
configurations with $a_{m} = a_{GW}$ are possible for values of $v_d < v_d(BP)$ it then follows that
GCs will in theory never be able produce binaries that only have the option of merging inside the cluster.
The relevant value of $a_{m}$ for GCs is then either $a_{ej}$ or $a_{tH}$. This of course could also be used to argue
why GCs have the properties they do, such as long time stability and no (visible) central massive BHs.
Indeed, several studies have shown that velocity dispersion do act as central parameter for distinguishing e.g. GCs from
NSCs with massive central BHs \citep[e.g.][]{2012ApJ...755...81M, 2016ApJ...831..187A}.
If some GCs have massive BHs in the range of $10^3 - 10^4 M_{\odot}$ in their center is still
the focus of both observational \citep[e.g.][]{2017Natur.542..203K}  and
theoretical work \citep[e.g.][]{2004ApJ...616..221G, 2015MNRAS.454.3150G, 2019arXiv190309659F, 2020MNRAS.491..113H},
and could provide insight into the formation of the super-massive BHs seen in most galactic centre \citep[e.g.][]{2012ApJ...755...81M, 2019MNRAS.486.5008A}.
Another feature linked to the BP is that clusters with $n \gtrsim n(BP)$
will (nearly) always produce and process binaries that undergo at least one IC due to the relative weak dependence on $v_d$ for the $n(GW,tH)$-boundary.
Regarding the dependence on $m$, one sees that the boundary quickly moves up for decreasing values of $m$, as
$n(BP) \propto m^{-13/8}$. This makes it increasingly difficult for 1G-objects with masses in the range $m \sim 1M_{\odot}$ to undergo more than 1 IC
within a Hubble time for astrophysical cluster values compared to $m \sim 30M_{\odot}$ 1G-objects, as clearly seen in Fig. \ref{fig:alim_fig}. Before we study this
in greater detail, we proceed below by exploring to which degree 3-body mergers and single-single (S-S) GW captures contribute to the in-cluster merger rate.

\subsubsection{3-body Mergers and Single-Single GW Captures}
\label{sec:Contribution from 3-body Mergers and S-S Captures}

Before moving on to how efficient a population of binaries is at producing a 2G population,
we here address the potential importance of including the in-cluster merger contribution from S-S GW captures and 3-body mergers.
As described in Sec. \ref{sec:Cluster Model}, `S-S GW captures' denote the process in which two initially unbound COs
become bound through the emission of GWs \citep[e.g.][]{2019arXiv190711231S}, where a `3-body merger' refers to COs merging during a
chaotic 3-body interaction \citep{2014ApJ...784...71S}.

We start by analyzing the contribution from 3-body mergers. For this, we
first estimate what part of the $(v_d,n)$-space the total integrated probability for producing a 3-body merger, $P_3$, is larger than the total probability for undergoing a
2-body merger, $P_2$. Following \cite{2018PhRvD..97j3014S}, the probability for a binary-single interaction to produce a 3-body merger
can be approximated by $p_3(a) \approx 2\mathscr{N}\left(\mathscr{R}_m/a\right)^{5/7}$, where $\mathscr{N}\approx 20$ denotes the number
of `temporary binary states' the chaotic triple interaction on average assembles during one interaction,
$\mathscr{R}_m$ is the Schwarzschild radius of a BH with mass $m$, and $a$ is the SMA of the initial target binary. The total probability for a 3-body merger to form during one
IC can now be found from integrating $p_3(a)$ from $a=a_{HB}$ to $a = a_m$, in the same way as we did for finding $P_2$ in Sec. \ref{sec:Overview...1}.
Following this approach, one finds that $P_3(a_m) \approx p_3(a_m) \times (7/(5\Delta))$, which can be written out in the following way,
\begin{equation}
P_3(a_m) \approx \left[ \frac{2^{12/7}7}{5\Delta} \frac{\mathscr{N}G^{5/7}}{c^{10/7}} \right] \times m^{5/7} a_m^{-5/7},
\label{eq:P3}
\end{equation}
where we have assumed that $p_3(a_m) \gg p_3(a_{HB})$ (Note here that we calculate these merger probabilities
separately, i.e., we do not take into account the potential interplay between merger channels, including the S-S GW capture channel).
From this we see that $P_3(a_m)/P_2(a_m) \propto (n/v_d)^{2/7}$. This indicates that 3-body mergers
will provide the greatest contribution relative to the 2-body mergers at high $n$ and low $v_d$, which is the regime where
$a_m = a_{ej}$, as seen on Fig. \ref{fig:alim_fig}. We therefore need to evaluate and compare
$P_2$ and $P_3$ for $a _m = a_{ej}$. Using Eq. \eqref{eq:P2}, Eq. \eqref{eq:P3}, and Eq. \eqref{eq:aej}, this first lead us to,
\begin{equation}
P_2(a_{ej}) \approx \left[ \frac{1}{B_{c}} \frac{f_{ed}^{10}}{G^3c^5} \right]^{2/7} \times \frac{v_d^{22/7}}{m^{6/7}n^{2/7}}
\label{eq:P2aej}
\end{equation}
and
\begin{equation}
P_3(a_{ej}) \approx \left[ \frac{42^{5/7}}{(5/63)} \frac{\mathscr{N} f_{ed}^{10/7}}{c^{10/7}}\right] \times v_{d}^{10/7}.
\end{equation}
Now setting these two expressions equal to each other one finds,
\begin{equation}
n(P_2,P_3) \approx \left[ \frac{(5/63)^{7/2}f_{ed}^5}{42^{5/2} \mathscr{N}^{7/2} B_c G^3} \right] \times \frac{v_{d}^{6}}{m^3},
\end{equation}
where $n(P_2,P_3)$ therefore represents the boundary in the $a_m = a_{ej}$ region for which $P_2 = P_3$.
This boundary is shown in Fig. \ref{fig:alim_fig} with the dotted line that encloses the black solid line hatched area.
In this area $P_3 > P_2$. As seen, for most systems, especially the one with relative low mass $m$ and moderate density $n$, 3-body mergers
will not dominate the total in-cluster merger probability. We will therefore in our analytical models throughout this paper omit
this contribution for simplicity and clarity. 

We now move on to the S-S GW capture population.
For this its more easy to compare merger rates, $\Gamma$, than probabilities.
In this case, the total rate of S-S GW capture mergers from a simple `$n \sigma v$' estimate is given by \citep{2019arXiv190711231S},
\begin{equation}
\Gamma_{ss} \approx \left[ \frac{4 \pi G^2}{c^{10/7}} \left(\frac{85\pi}{24\sqrt{2}}\right)^{2/7} \right] \times \frac{N_sm^2}{v_d^{18/7}},
\label{eq:Gss}
\end{equation}
where $N_s$ is the total number of single BHs.
Note that we have here assumed that all the single BHs, $N_s$, are distributed uniformly
according to our model of a constant $v_d,n$; however, in reality, the single BHs naturally distribute according to some density and velocity
profile. As a result, the real GW capture rate is generally smaller than the one presented in the above Eq. \eqref{eq:Gss},
as further discussed in \cite{2019arXiv190711231S}. Regarding the merger rate from our considered binary-single interactions, one finds that this can be approximated by,
\begin{equation}
\Gamma_{23} \approx \frac{N_b(P_2(a_{m}) + P_3(a_{m}))}{{\tau}_m},
\label{eq:Gbs}
\end{equation}
where $P_2 + P_3 \leq 1$ is the total number (probability) of 2-body and 3-body mergers forming during 1 IC, $\tau_m$ is the time it takes
for undergoing 1 IC (see Eq. \eqref{eq:tauaf}), and $N_b$ is the number of CO binaries in the cluster that contributes to the merger rate.
We have here included the 3-body mergers, as it turns out that
the S-S GW captures only significantly contribute for low $v_d$ and high $n$, exactly where the 3-body mergers also contribute.
This is seen in Fig. \ref{fig:alim_fig}, where the black dotted line inclosing the black dotted area is where $\Gamma_{ss} = \Gamma_{23}$
for binary fraction $N_b/N_s = 0.05$. The S-S GW captures are therefore not expected to provide a significant contributing in the regions we are interested in.

To conclude, we have here shown and argued that neither the 3-body mergers nor the S-S GW capture mergers
contribute significantly to the in-cluster merger rate. We therefore only consider the 2-body merger contribution in the rest of this paper.

\subsubsection{Interaction Cycles and In-cluster Mergers}
\label{sec:Burning Cycles and Merger Efficiency}

\begin{figure}
\centering
\includegraphics[width=\columnwidth]{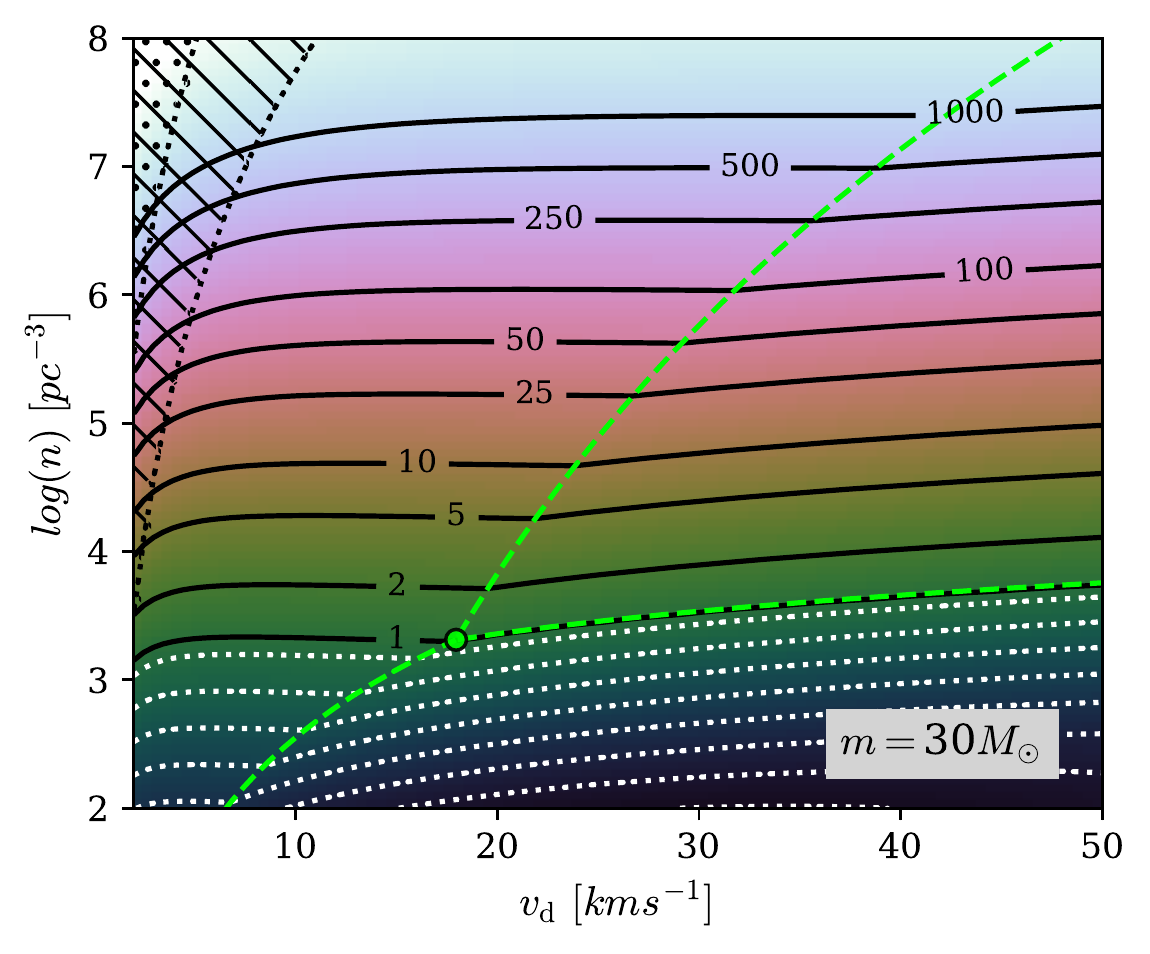}
\includegraphics[width=\columnwidth]{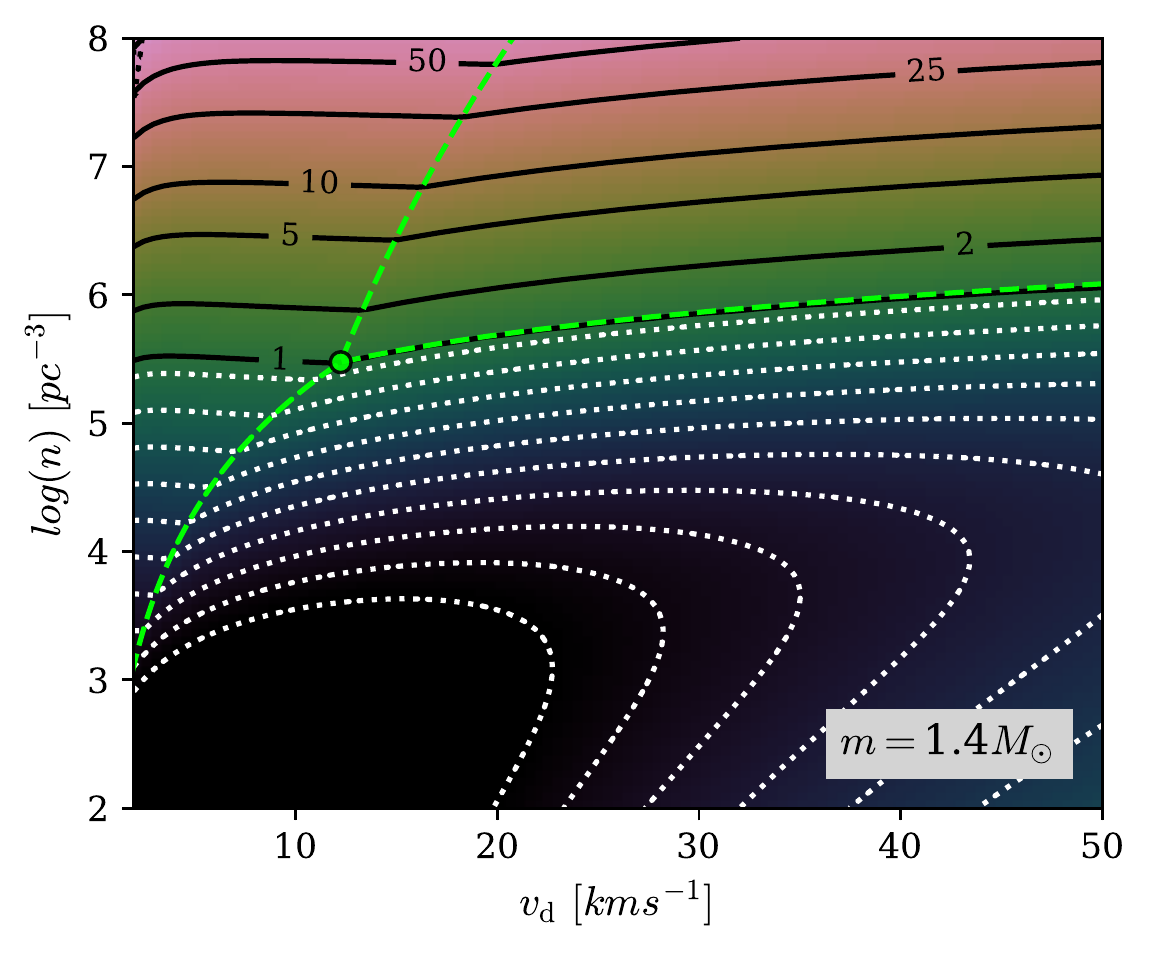}
\caption{Similar to Fig. \ref{fig:alim_fig}, but here the {\it black solid lines} show the corresponding number of in-cluster
GW mergers that form in a Hubble time per binary, $N'_{M}(t_H)$. The {\it white dotted} contour lines show where
$N'_{M}(t_H) < 1$, and highlights therefore the area for which the system is not effective in growing a 2G population.
For this figure we have assumed that $N'_{M}(t_H) \approx N_c(t_H) \times P_{M}$, where $N_c(t_H)$ is the number of ICs
a binary can undergo in a Hubble time, and $P_{M} = P_2 + P_3$ is the probability for a binary to undergo an in-cluster GW merger
during one IC. These two quantities are also shown separately in Fig. \ref{fig:alim_fig}. As seen in the figure above,
the number $N'_{M}(t_H)$ is almost independent of $v_d$, but highly sensitive to especially $m$.
}
\label{fig:Nmbin}
\end{figure}

The number of in-cluster GW mergers that can be produced over a Hubble time per 1G-1G binary, here denoted by $N'_{M}(t_H)$,
serves as an approximate measure for how efficient a given cluster is at growing a 2G population. At this stage we
approximate this number by the following product,
\begin{equation}
N'_{M}(t_H) \approx N_c(t_H) \times P_{M},
\label{eq:NIMtH}
\end{equation}
where $N_c(t_H) = t_H/{\tau}_m$ is the number of ICs a cluster can run through in a Hubble time, i.e.
the number of binaries the cluster can process in time $t_H$, and $P_{M}$ is the probability for an in-cluster GW merger to form during one IC.
In Fig. \ref{fig:alim_fig} is shown with {\it orange dashed lines} and {\it red solid lines} the contours of $N_c(t_H)$ and $P_{M}$, respectively,
where for $P_{M}$ we have here included the probability for 3-body mergers, i.e. $P_{M} = P_2(a_m) + P_3(a_m)$. 
In short, our procedure for estimating $N_c(t_H)$ and $P_{M}$ at a given point $(v_d, n)$, is first to calculate $a_m$
from Eq. \eqref{eq:amin}, after which we use Eq. \eqref{eq:tauaf} to find $N_c(t_H) = t_H/\tau_m$, and Eq. \eqref{eq:P2} and Eq. \eqref{eq:P3}
to find $P_{M} = P_2(a_m) + P_3(a_m)$.

As seen in Fig. \ref{fig:alim_fig}, for decreasing values of $v_d$ the probability $P_{M}$ decreases, which also follows from 
Eq. \eqref{eq:P2aej} where $P_2 \propto v_d^{22/7}$, in contrast to the number of ICs, $N_c(t_H)$, that instead increases. Therefore, one can easily have a
cluster with binaries where burning is efficient, i.e. where $N_c(t_H) \gg 1$, but at the same time with a probability for merging during individual
ICs is low, i.e. with $P_{M} \ll 1$, and vice versa. How these two quantities `balance out' is clear in Fig. \ref{fig:Nmbin}, which shows in black solid lines
$N'_{M}(t_H)$ from Eq. \eqref{eq:NIMtH}. Surprisingly, the large changes in both $P_{M}$ and $N_c(t_H)$
as $v_d$ is varied almost cancel out, and $N'_{M}(t_H)$ is as a result almost flat across $v_d$.
To study this behavior further, we can write out $N'_{M}(t_H)$ in the region relevant for GC systems where $P_2 \gg P_3$ and $a_m = a_{ej}$
using Eq. \eqref{eq:NIMtH}, Eq. \eqref{eq:tauaf} evaluated at $a_m = a_{ej}$, and the expression for $P_{2}(a_{ej})$ given by Eq. \eqref{eq:P2aej}, from which one finds, 
\begin{align}
N'_{M}(t_H) & \approx t_H \left[ \left(\frac{4\pi}{63}\right)^{7/2} \frac{G^4f_{ed}^{3}}{{B_c}{c^5}} \right]^{2/7} \times n^{5/7}m^{8/7}v_d^{1/7} \\
                & \approx 0.5 \left(\frac{n}{10^{5}pc^{-3}}\right)^{5/7} \left(\frac{m}{1.4M_{\odot}}\right)^{8/7} \left(\frac{v_d}{10\ kms^{-1}}\right)^{1/7},
\label{eq:NIMtH_aej}
\end{align}
where in the last line we have inserted values relevant for NS-NS mergers.
This confirms the results we see in Fig. \ref{fig:Nmbin}, namely that $N'_{M}(t_H)$ only depends weakly on $v_d$ as $N'_{M}(t_H) \propto v_d^{1/7}$.
As a result, all systems with $n \gtrsim n(BP)$ will to leading order have $N'_{M}(t_H) \gtrsim 1$. From this follows that if the number of CO binaries is constant
in time at a value $N_b$, then the number of in-cluster mergers for $n \gtrsim n(BP)$ will be $\gtrsim N_b$ after a Hubble time.
For example, for our $m = 30 M_{\odot}$ case shown in the upper panel of Fig. \ref{fig:Nmbin}, the number of in-cluster mergers
over a Hubble time per binary is of order $10$ for $\log{n} \approx 4 \sim 5\ pc^{-3}$. If the number of BBHs in the cluster
at any given time is a few, say $\sim 5$, then our model predicts that the total number
of in-cluster mergers forming over a Hubble time is $\sim 5 \times 10 = 50$. Although this number of course fluctuates from cluster to cluster,
we note that this number is consistent with what is found using numerical simulations (see \cite{2019PhRvD.100d3027R}, where $48$ in-cluster mergers
were reported for their example in Sec. IV.A). More generally, $N'_{M}(t_H)$ provides an upper limit on the number of available 2G objects
after a Hubble time produced per cluster binary, as only a small fraction of the in-cluster mergers, i.e. 2G objects, are actually retained by the cluster \cite{2019PhRvD.100d3027R}.
The remaining are either kicked out immediately as a result of GW kicks, or later dynamically through e.g. a binary-single interaction.
Considering the $m=1.4 M_{\odot}$ case, we see both from Eq. \eqref{eq:NIMtH_aej} and Fig. \ref{fig:Nmbin} that $N'_{M}(t_H)$ is only
$\gtrsim 1$ for $\log{n} \gtrsim 5-6\ pc^{-3}$, which is a very high density threshold for astrophysical standards. This provides a clear hint that clusters hosting only NS-NS binaries are not likely to be effective in turning its population into
a sizable 2G population, i.e. in populating the LMG, unless the binary fraction initially is relatively high. 

Lastly, in relation to the probability of observing a possible 2G-population from a cluster, what
matters is not only the number of 2G-objects produced, but also how many of these that are present in the
cluster compared  to the number of remaining 1G objects. As described back in Sec. \ref{sec:Overview...1},
a single IC will on average give rise to $N^{s}_{ej} + 2 \sim 6$ ejected 1G-objects (if in-cluster mergers and 2G objects are ignored), which naturally leads to a gradual reduction of this population over time.
In our model considered so far, the number of in-cluster GW mergers relative to the number of (remaining) 1G-objects after time $t$ is therefore approximately,
\begin{align}
\frac{N'_{M}(t)}{N_1(t)}    & \approx \frac{N_{b}N'_{M}(t)}{N_{1}(0) - N_{b}N_{c}(N^{s}_{ej} + 2)}\\
                                & \approx \frac{N'_{M}(t)}{f^{-1}_{b}(0) - N'_{ej}(t)},
\label{eq:NIMoN1t}
\end{align}
where we have assumed that $N_b$ remains constant, $N_{1}(0)$ denotes the initial
number of $1G$-objects, $f_b(0) = N_{b}/N_{1}(0)$, and $N'_{ej}(t)$ denotes the total number
of $1G$-objects ejected after time $t$ per binary. We will explore this ratio and others in the sections below.

\section{Populating the Black Hole Mass Gaps: Time Evolving Cluster Model}
\label{sec:Populating the Black Hole Mass Gaps: Time Evolving Cluster Model}

In this section we develop a simple time dependent cluster model to study the evolution of
both 1G and 2G objects as a function of time. As further described in the following sections,
in this model we take into account both binary and single dynamical ejections, and in particular the growth
of 2G objects as a result of in-cluster 1G-1G mergers. We (still) assume the cluster is described
by a fixed set $v_d, n$, and all objects have the same mass $m$, which of course is a simplification of a real cluster.
This in turn however enables us to put forward simple, general, and informative statements, based solely on
characteristic mass, length, and time scales.

In the first section below we derive a set of evolution equations for
$N_1$ and $N_2$, where $N_i$ here denotes the number of objects of type `$i$'. In Sec. \ref{sec:Nevol_Results} we solve these equations from which we put upper limits on the ratio $N_2/N_1$, illustrated for $m = 1.4M_{\odot}$ (2G in the LMG)
and $m=30M_{\odot}$ (2G in the UMG), for a grid of cluster systems described by $v_d,n$.

\subsection{Evolution Equations}
\label{sec:Evolution Equations}

\begin{figure}
\centering
\includegraphics[width=\columnwidth]{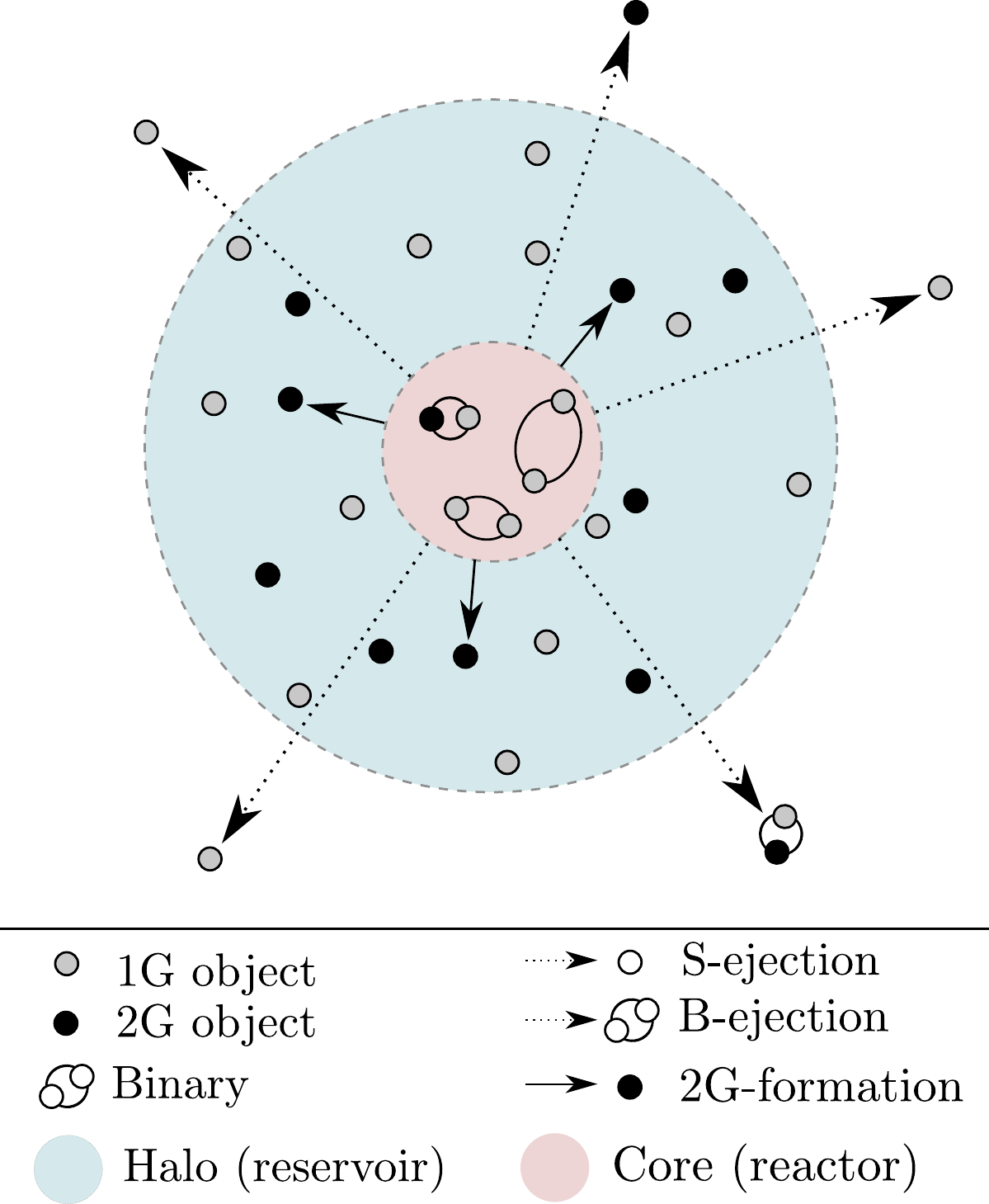}
\caption{Illustration of the cluster model we use to study the formation of 2G-objects ({\it black dots}) through successive in-cluster GW mergers
of 1G-objects ({\it grey dots}). The cluster is composed of two parts, an inner core ({\it pink region}) and an outer halo ({\it blue region}). Inside the
core there is $N_b$ binaries that interact with the flow of objects coming in from the outer halo. The resulting binary-single interactions
are modeled using our $\delta$-model described in Sec. \ref{sec:Cluster Model}, which leads to both dynamical
ejections of binaries ({\it B-ejection}) and singles ({\it S-ejection}),
and the production of 2G-objects ({\it 2G-formation}) through in-cluster GW mergers of 1G-1G binaries.
The number of 2G-objects compared to the number of 1G-objects, i.e. $N_2/N_1$, after a Hubble time
provides a rough estimate for how likely it is to observe a GW source from a binary merger that includes at least one 2G-object.
We study the evolution of $N_2/N_1$ in Sec. \ref{sec:Populating the Black Hole Mass Gaps: Time Evolving Cluster Model},
and comment on the implied possibility for populating the UMG and the LMG in Sec. \ref{sec:Upper Limits on 2G Objects}.}
\label{fig:cluster_ill}
\end{figure}

We consider a cluster described by a constant $v_d,n$ that initially has a population of $N_1(0)$ 1G objects all with equal mass $m$.
In this cluster there is a (time-dependent) population of binaries that interact with the surrounding single population,
which give rise to dynamical ejections, exchanges, and in-cluster mergers. The absolute and relative number of $N_1$ (1G) and $N_2$ (2G) objects
therefore changes over time through various dynamical mechanisms, that all depend on $v_d, n, m$. This configuration is described and illustrated in
Fig. \ref{fig:cluster_ill}. The question is, for what initial conditions of $v_d, n, m$ is the system able to produce a sizable population of 2G objects after a Hubble time?
To answer this question, we start by writing out the following set of differential equations that we take to represent the evolution of
$N_1$ and $N_2$,
\begin{align}
    \dot{N}_{1}    = &-\left( \dot{N}_1^{ej} + \dot{N}_{21}^{ej} + 2\dot{N}_{11}^{ej} \right) - \left( \dot{N}_{21}^{M} + 2\dot{N}_{11}^{M} \right)  \nonumber\\
    \dot{N}_{2}    = &-\left( \dot{N}_2^{ej} + \dot{N}_{21}^{ej} + 2\dot{N}_{22}^{ej} \right) - \left( \dot{N}_{21}^{M} + 2\dot{N}_{22}^{M}\right)  \nonumber \\
     			&+ \left(R_{11}^{M}\dot{N}_{11}^{M} \right),
\label{eq:N1N2_gen1}
\end{align}
where $N^{ej}_{i}$ is the number of objects of type `$i$' (1G or 2G) that are ejected as singles, $N^{ej}_{ij}$ is the number of ejected binaries consisting of object types `$\{ij\}$',
$N^{M}_{ij}$ is the number of $\{ij\}$-binaries merging in-side the cluster, and $R_{11}^{M}$ is the retention fraction of 1G-1G mergers. As seen, both `ejections' (single and binary)
and `in-cluster mergers' all act as `sink terms', except as for the term $\propto {N}_{11}^{M}$ that serves as the 2G `source term'.
As we are studying the process of growing a 2G population through in-cluster GW mergers during successive ICs, we
restrict our self in the following to describe systems that are able to undergo $N_c \gg 1$.
Therefore, instead of evolving the above equations over e.g. individual interaction steps `$k$', or time,
we evolve them over the number of ICs, $N_c$. The `dot' over each $N$ refers therefore to the change per IC.

The relevant terms for writing out our evolution equations from above can be written as,
\begin{alignat}{3}
&\dot{N}_{i}^{ej}  &&\approx [N_b \bar{P}_M] p_{i}^{ej}N_{s}^{ej},\\
&\dot{N}_{ij}^{ej} &&\approx [N_b \bar{P}_M] p_{ij}^{ej},\\
&\dot{N}_{ij}^{M} &&\approx [N_b {P}_M] p_{ij}^{b}, 
\end{alignat}
where $N_{b}$ is the number of binaries,  ${P}_M$ ($\bar{P}_M = 1-{P}_M$) is the integrated probability that a given binary does (not) merge during a single IC,
$p_{i}^{ej}$ is the probability that object `$i$' is ejected after a binary-single interaction, $N_{s}^{ej}$ is the total number of singles per binary ejected during
one IC, $p_{ij}^{ej}$ is the probability that binary-`$\{ij\}$' is ejected after a binary-single interaction, and $p_{ij}^{b}$ is the probability that $\{ij\}$ is in a binary at
a random hardening step `$k$'. These terms can be further expanded as,
\begin{alignat}{4}
& {p}_{2}^{ej}  		&& \approx p_{2}^{i} p_{112}^{es} [1+B],\ \ \ \ 	&& {p}_{1}^{ej} 		&&\approx 1-{p}_{2}^{ej} \\
& {p}_{21}^{ej}		&& \approx p_{2}^{i} p_{211}^{es} [1+B],\ \ \ \ 	&& {p}_{11}^{ej}	&&\approx 1-{p}_{21}^{ej} \\
& {p}_{21}^{b}   	&& \approx p_{2}^{i}B, \ \ \ \ 				&& {p}_{11}^{b}		&&\approx 1-{p}_{21}^{b} \\
& {p}_{211}^{es}	&& \approx 2w/3, \ \ \ \ 					&& {p}_{112}^{es} 	&&\approx 1-{p}_{211}^{es}\\
& p_{2}^{i}			&& \approx N_2 F/(N_1 + N_2), \ \ \ \			&& p_{1}^{i} 		&&\approx 1-p_{2}^{i}.
\end{alignat}
where $p_{2}^{i}$ is the probability that object type `$2$' (2G) is the incoming single object in a binary-single interaction at
hardening step `$k$', $p_{ijk}^{es}$ is the probability that a given binary-single interaction results in an endstate where $\{ij\}$ is a
binary and `$k$' leaves as single, and $B = 2Fw/(3-2w)$. The factor $F$ is introduced to quantify the probability `enhancement' of
a 2G-object to interact with a binary compared to a 1G-object. For example, the enhancement factor from standard gravitational focusing of having
a 2G-object to interact with a binary compared to a 1G-object is $F = (1+1+2)/(1+1+1) = 4/3$.
Similarly, $w$ describes the `enhanced probability' that the outcome of a binary-single interaction involving a 2G-object is
$\{121\}$, i.e. where `$\{12\}$' is a binary and `$1$' is ejected as single. For this set of equations
we have made four central assumptions:  (1) All binary-single interactions involving objects $\{ijk\}$ have the same
outcome distributions irrespective of the initial configuration. (2) The probability to have interactions with $> 1$ 2G-object is $= 0$,
which follows from our considered limit of $N_2 \ll N_1$. (3) Dynamical single and binary ejections associated with a given interacting binary are
only $>0$ if the binary in question does not merge before concluding its IC. (4) All interactions and ICs follow our `$\delta$-model' illustrated in Fig. \ref{fig:ill_int}.
Now using these equations we can rewrite our evolution equations given by Eq. \eqref{eq:N1N2_gen1} as follows,
\begin{align}
    \dot{N}_{1}    &=  N_{b} \times \left[+p^{i}_{2}\left(A - P_{M}(A-B)\right) -  \left({N}_{t}^{ej} - P_M {N}_{s}^{ej}\right) \right] \nonumber\\
    \dot{N}_{2}    &=  N_{b} \times \left[- p^{i}_{2}\left(A - P_{M}(A-B)\right) + \left(p_{11}^{b} P_{M} R_{11}^{M}\right) \right],
\label{eq:N1N2_ABform}
\end{align}
where ${N}_{t}^{ej} = 2+{N}_{s}^{ej}$ is here the total number of ejected objects over 1 IC,
and $A = [1+B]\left(p_{112}^{es} N_{s}^{ej} + p_{211}^{es}\right)$.

To summarize, our presented evolution equations given Eq. \eqref{eq:N1N2_gen1} are completely general, and shows simply what characteristic sink and 
source terms that are relevant for our problem. Other terms, such as strong binary-binary interactions \citep{2019ApJ...871...91Z}, and
weak few-body scatterings \citep{2019MNRAS.487.5630H, 2019PhRvD.100d3010S, 2019MNRAS.488.5192H, 2020MNRAS.494..850H},
or more general mass-ratio dependent terms and corresponding GW kick prescriptions can be included, but this is beyond this paper.
The resulting terms shown in Eq. \eqref{eq:N1N2_ABform} follow directly from simple combinatorics, and are constructed by
calculating the (time dependent) probability for 1G- and 2G-objects to interact and exchange into the interactions states
shown in Fig. \ref{fig:ill_int}, folded with the probability for dynamical ejections and in-cluster mergers during each IC.
In the following sections we consider solutions to this coupled set of equations, from which we especially find a closed form
solution to the upper limit on $N_2/N_1$ as a function of time.

\subsection{Results}
\label{sec:Nevol_Results}

In the first section below, we study the evolution of $N_1$ and $N_2$ for two different cluster models, denoted $cA$ and $cB$,
using the general set of evolution equations presented in the above Sec. \ref{sec:Evolution Equations}.
In the second section, we use these results to study the upper limit on the ratio $N_2/N_1$ evaluated at present day, i.e. at $t = t_H$,
for a grid of $v_d,n$ cluster systems.

\subsubsection{Time-Evolving Populations}
\label{sec:Time-Evolving Populations}

We study the evolution of $N_1$ and $N_2$ using Eq. \eqref{eq:N1N2_ABform} for two distinct cases, $cA$ and $cB$.
These two cases are described in the following.

\begin{figure}
\centering
\includegraphics[width=\columnwidth]{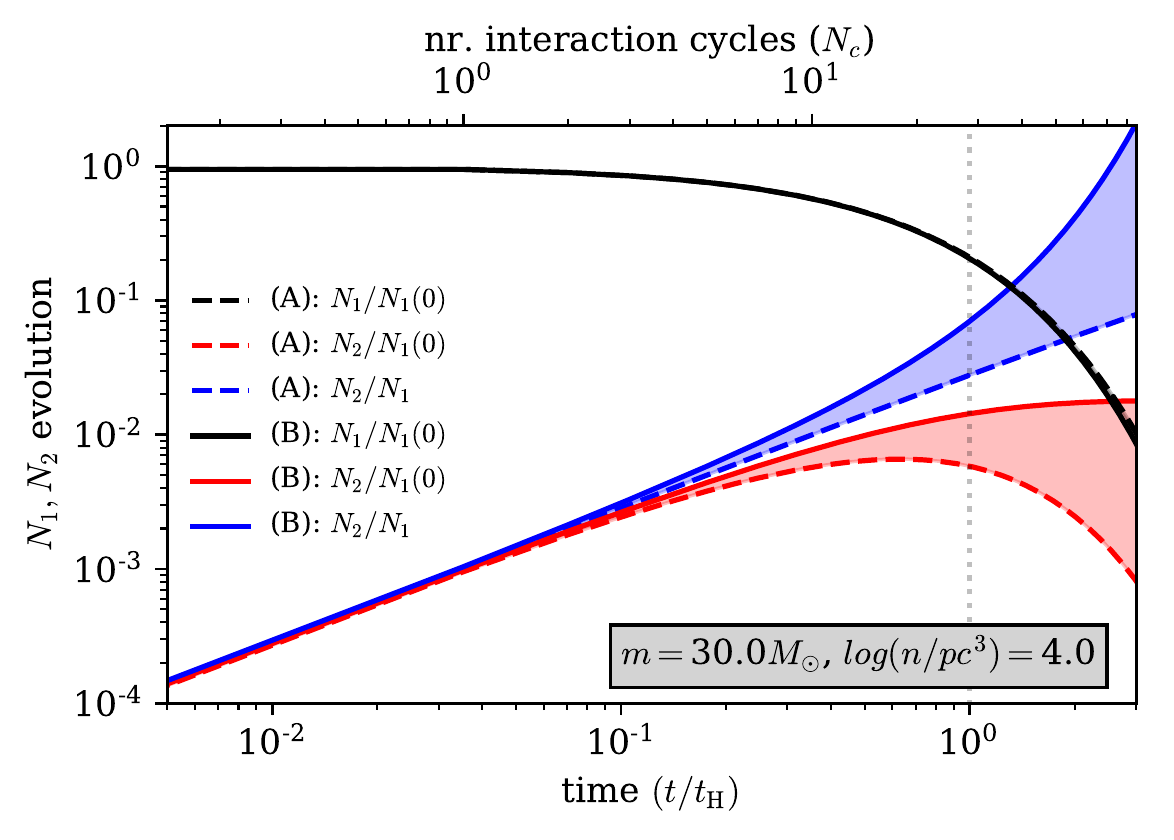}
\includegraphics[width=\columnwidth]{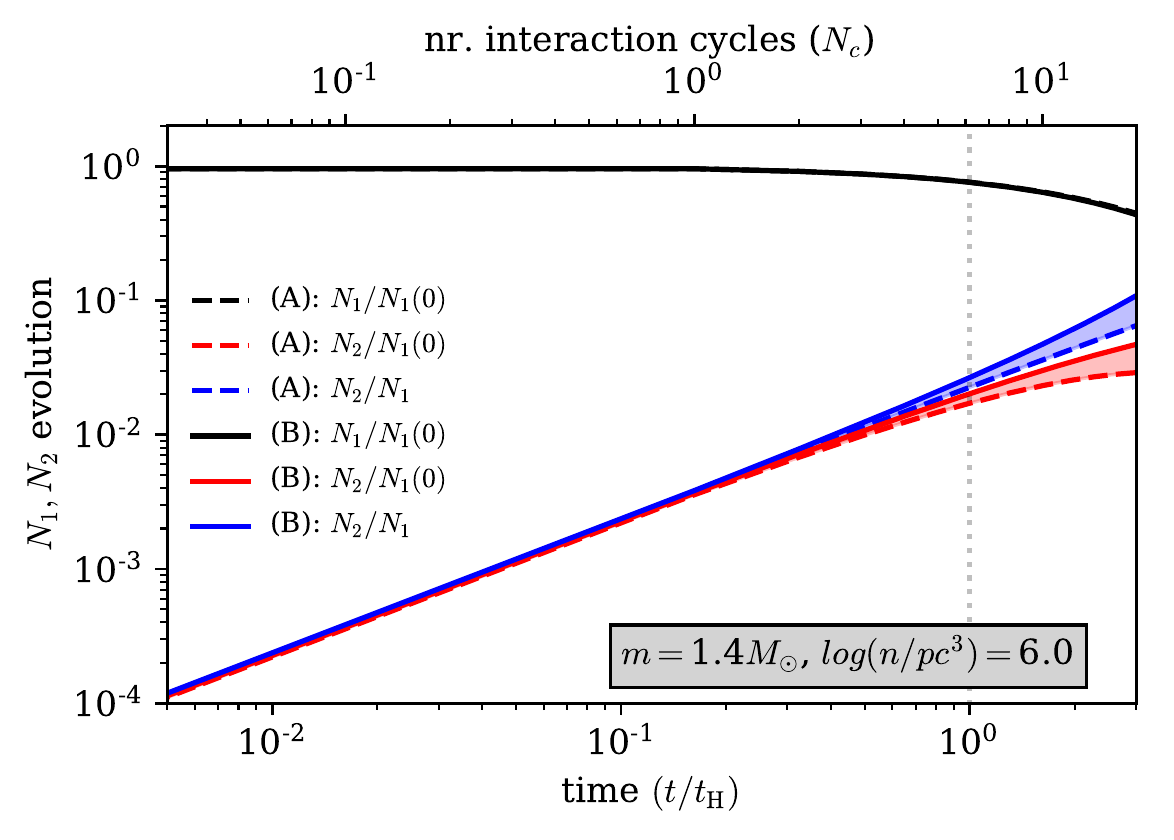}
\caption{Re-scaled evolution of 1G-objects ({\it black lines}), $N_1/N_1(0)$, 2G-objects ({\it red lines}), $N_2/N_1(0)$, and their number ratio ({\it blue lines}), $N_2/N_1$,
as a function of time ({\it lower x-axis}), $t/t_H$, and corresponding number of ICs ({\it upper x-axis}), $N_c = t/\tau_m$. The {\it dashed lines} and {\it solid lines}
show the solution from our cases $cA$ and $cB$ described in Sec. \ref{sec:Nevol_Results}, respectively.
For both plots we have assumed that $f_b = 0.01$, $N_{t}^{ej} = 6$, $v_d = 10\ kms^{-1}$, and $R_{m}^{11} = 1$,
where the {\it upper plot} shows results for $m=30M_{\odot}$, and the {\it lower plot} for $m = 1.4M_{\odot}$.
In the $m=30M_{\odot}$ case the ratio $N_2/N_1$ approaches the $10\%$-level at $t/t_H \sim 1$ for our chosen parameters,
which indicates that populating the UMG through in-cluster 1G-1G GW mergers seems possible. This is in contrast to the
$1.4M_{\odot}$ case, where $n$ needs to take the relative high value of $n \sim 10^{6}pc^{-3}$ to even reach the $1\%$-level.
This is further discussed in Sec. \ref{sec:Nevol_Results} and Sec. \ref{sec:Upper Limits on 2G Objects}.}
\label{fig:Nevolv}
\end{figure}

{\it Case `cA':} In this case we assume the weight factors $F=1$ and $w=1$, i.e., we keep track of the growing population of 2G-objects,
but assume that in all dynamical aspects a 2G object is indistinguishable from a 1G object. We are therefore
able to explore the effect from pure `combinatorics' arising from the growing population of 2G-objects that are free to exchange, merge, and being
ejected in the same way as the 1G objects. Using Eq. \eqref{eq:N1N2_ABform} with $F = 1, w = 1$
the evolution equations are in this case given by,
\begin{align}
    \dot{N}_{1}/N_{b}    	\approx	&-p_{1}^{i} \left(N_{t}^{ej} - P_M N_{s}^{ej}\right) \nonumber\\
    \dot{N}_{2}/N_{b}    	\approx	&-p_{2}^{i} \left(N_{t}^{ej} - P_M N_{s}^{ej}\right) + P_M R^{M}_{11} \left(1 - 2p_{2}^{i}\right),
\label{eq:N1N2cA}
\end{align}
where we have used that under these assumptions $A = N_{t}^{ej}$ and $A - B = N_{s}^{ej}$.
In this case the number of 2G-objects compared to 1G-objects present in the cluster after a Hubble time represents approximately
a lower limit, as in `reality' a higher number of 2G-objects will be left in the cluster due to their higher mass \citep[e.g.][]{Sigurdsson:1993jz}.

{\it Case `cB':} In this case we assume that $p_{2}^{i} = 0$ and $R_{11}^{M} = 1$,
i.e. that the 2G-objects are not participating in any interactions, and that the 1G-objects as a result
dynamically evolve through interactions, merger, and ejections completely independent of the 2G-objects. As a result, the number of 2G-objects we here find
after time $t_{H}$ represents the highest number possible, and the 1G-population will also decrease to its lowest possible value. This case
therefore represents the upper limit on how many 2G-objects one can keep in a cluster after time $t_H$ compared to the 1G-population.
The evolution equations are in this case given by Eq. \eqref{eq:N1N2_ABform} with $p_{2}^{i} = 0$ and $R_{11}^{M} = 1$,
\begin{align}
    \dot{N}_{1}/N_{b}    &\approx  -   \left(N_{t}^{ej} - P_M N_{s}^{ej}\right) \nonumber\\
    \dot{N}_{2}/N_{b}    &\approx  +  \left(P_M \right).
    \label{N1N2cB}
\end{align}
This set of equations have a particular simple and interesting set of analytical solutions that we now explore before moving on.
For this, we start by rewriting the above equations into a more general form to shorten the notations:
$\dot{N}_{1} = - \alpha N_{b}$,  $\dot{N}_{2} = \beta N_{b}$, where we have defined,
\begin{align}
   \alpha	& = N_{t}^{ej} - P_M N_{s}^{ej} \nonumber\\
   \beta	& = P_M.
\end{align}
To proceed, we now consider a specific model where the binary fraction stays constant such that $N_b = f_b \times N_1$.
In this case, the solution to the above set of equations is easily found from simple integrations, from which it follows,
\begin{align}
    N_1	&=  N_1(0) \times \exp{(-\alpha f_b N_c)} \nonumber\\
    N_2	&=  N_1(0) \times ({\beta}/{\alpha}) \left[ 1 - \exp{(-\alpha f_b N_c)} \right],
    \label{eq:N1N2_sol_ab}
\end{align}
where $N_1(0)$ is the initial number of 1G-objects, and $N_c = t/\tau_m$ is the number of ICs after time $t$.
If we first consider the solution to $N_1$, we see that the population of 1G-objects `decays' over time as if the cluster represents a giant `radioactive nuclei'
with decay time $t_{cd}$, given by
\begin{equation}
   t_{cd}	\approx \frac{\tau_m}{{\alpha}f_b},
\end{equation}
where the time for undergoing one IC, $\tau_m$, is given by Eq. \eqref{eq:tauaf}. For example, for $a_m = a_{ej}$ the decay time is
$t_{cd} \propto v_{d}^{3}/(n m^2 f_{b})$, where we have used Eq. \eqref{eq:aej}.
One consequence of this model is that the decay rate, and thereby the number of 1G-objects $N_1$ after a Hubble time, depends exponentially on the binary fraction $f_b$.
The binary fraction is at the moment unknown observationally, but numerical simulations of GCs using Monte-Carlo techniques have shown that it very likely
stays constant with only small scatter around $1-5\%$ (see e.g. Fig. 2 in \citealt{2019arXiv190711231S}).
As a result, a significant fraction of present day GCs likely have many of their 1G-objects left in their core, where the remaining fraction have lost its BHs through
binary-single `evaporation'. This `evaporation effect' will lead to a characteristic change in BBH merger rates as a function of redshift,
similar to what is found for the set of GCs that `evaporates' through tidal heating or direct tidal disruptions \citep[e.g.][]{2018PhRvL.121p1103F}.
Considering now $N_{2}$ for our model, we see that at early times $N_2 \approx N_{1}(0)f_{b} P_{M} N_c$, where we have used that $\exp(-ax) \approx 1 - ax$. This
is expected, as this simply equals the number of mergers per IC evaluated for the initial $N_1(0)$ population
($N_{1}(0)f_{b} P_{M}$) times the number of ICs ($N_c$). Note that this is similar to Eq. \eqref{eq:NIMtH}, where we studied
how effective a population consisting of a single binary (`$1 = N_{1}(0)f_{b}$') is at growing a 2G-population.
As $N_c$ increases towards infinity, the $N_2$ population reaches a maximum `freeze-out value', $max(N_2)$, given by
\begin{equation}
   max(N_2) = N_{1}(0) ({\beta}/{\alpha}),\ N_c \rightarrow \infty,
   \label{eq:maxN2}
\end{equation}
which interestingly do not depend on the binary fraction, although how fast $N_2$ reaches $max(N_2)$ does.
As seen, within a factor of unity, $max(N_2)$ its simply given by the total number of binary mergers one would get if one turned the initial
$N_1(0)$ population into a total of $N_1(0)/2$ binaries.
Finally, if we now consider the number of 2G-objects relative to 1G-objects, one finds using Eq. \eqref{eq:N1N2_sol_ab} that
\begin{equation}
   {N_2}/{N_1}    =  ({\beta}/{\alpha}) \left[ \exp{(\alpha f_b N_c)} - 1 \right].
   \label{eq:N2oN1}
\end{equation}
We see here that this ratio always increases, i.e., in this case there is no `freeze-out' value. This of course originates from that $N_1$ keeps decreasing,
whereas $N_2$ keeps increasing until it asymptotically reaches its value $max(N_2)$.
Considering the limit where $N_2/N_1 = 1$, we can solve for the corresponding characteristic $N_c$ scale, denoted here by $N_{c}^{2E1}$ ,
\begin{equation}
  N_{c}^{2E1} = \frac{ln\left(1 + \alpha/\beta \right)}{\alpha f_b},
  \label{eq:Nc21}
\end{equation}
which equals the number of IC cycles, or time $t_{c}^{2E1} \approx N_{c}^{2E1} \times \tau_m$,
it takes for $N_2$ to be similar to $N_1$. Comparing $t_{c}^{2E1}$ with $t_H$ provides a rough estimate for when a system
is effective in growing a sizable 2G-population within a Hubble time.
We will study the ratio $N_2/N_1$ from $cB$ in greater detail in Sec. \ref{sec:Upper Limits on 2G Objects} below.

The evolution of $N_1$ and $N_2$ for case $cA$ and $cB$ as a function of time is shown in Fig. \ref{fig:Nevolv}
assuming the binary fraction stays constant at $f_b = 0.01$, $N_{t}^{ej} = 6$, $N_{s}^{ej} = 4$, $v_d = 10\ kms^{-1}$, and $R_{M}^{11} = 1$.
Note here that in the upper plot where $m = 30M_{\odot}$ the density is $n=10^{4}\ pc^{-3}$,
whereas in the lower plot for $m=1.4 M_{\odot}$ the density is instead $n = 10^{6}\ pc^{-3}$, as this is around the threshold for
when $N_c \gg 1$ (see Fig. \ref{fig:alim_fig}).
Starting with $cA$, we see in the $m=30M_{\odot}$ case how the 2G-population first grows steadily up to a given point
just before $t=t_H$, after which it starts decreasing. This decrease is sourced by the binary and single ejection sink terms from Eq. \eqref{eq:N1N2cA}.
Considering now $N_2/N_1$, we see that at $t=t_H$ this ratio is (only) at the $1\%$-level. We therefore expect $N_2/N_1$
to be of that order or greater for these cluster values, depending on the retention fraction $R_{M}^{11}$.
The same characteristics are true for the $m=1.4M_{\odot}$ case, but to reach a value for $N_2/N_1$ of a few percent, we see that
$n$ in this case has to be of order $10^{6}\ pc^{-3}$, which is much higher than what is found in most astrophysical systems.
Considering now $cB$, it is seen for $m=30M_{\odot}$ that $N_1$ decays exponentially, whereas $N_2$ steadily levels off at its `freeze-out value' given
by Eq. \eqref{eq:maxN2}. The characteristic time given by Eq. \eqref{eq:Nc21} for which $N_2 = N_1$
is only $2 \sim 3$ times $t_H$, and as a result, the ratio $N_2/N_1$ approaches here the $10\%$-level at $t_H$.
This indicates that $\gtrsim 30M_{\odot}$ COs are able to reach interesting limits when it comes to populating the UMG,
whereas in the $\sim 1M_{\odot}$ CO case, it seems very difficult to undergo enough in-cluster mergers to populate the
LMG. We will study this in greater detail in the section below.

\subsubsection{Upper Limits on 2G-Objects}
\label{sec:Upper Limits on 2G Objects}

\begin{figure}
\centering
\includegraphics[width=\columnwidth]{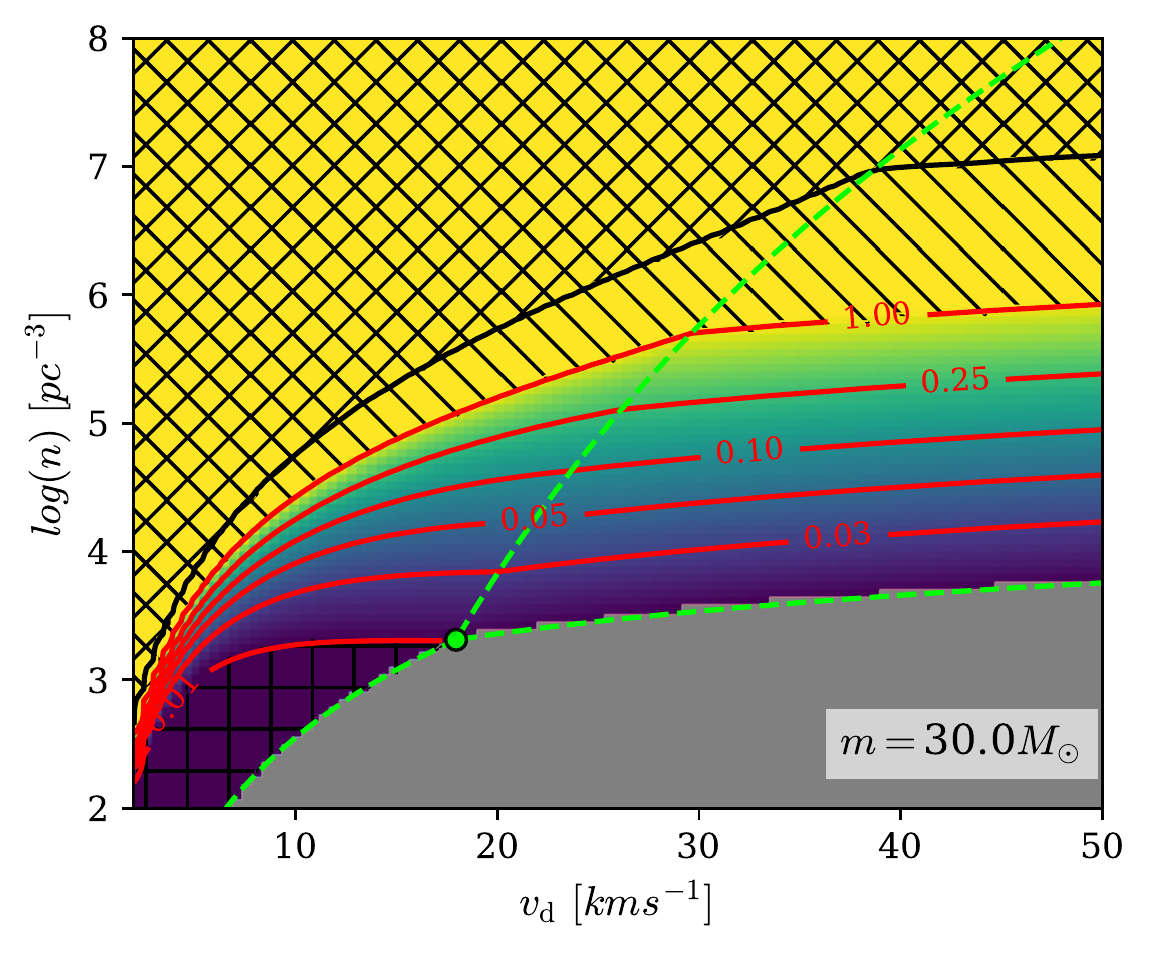}
\includegraphics[width=\columnwidth]{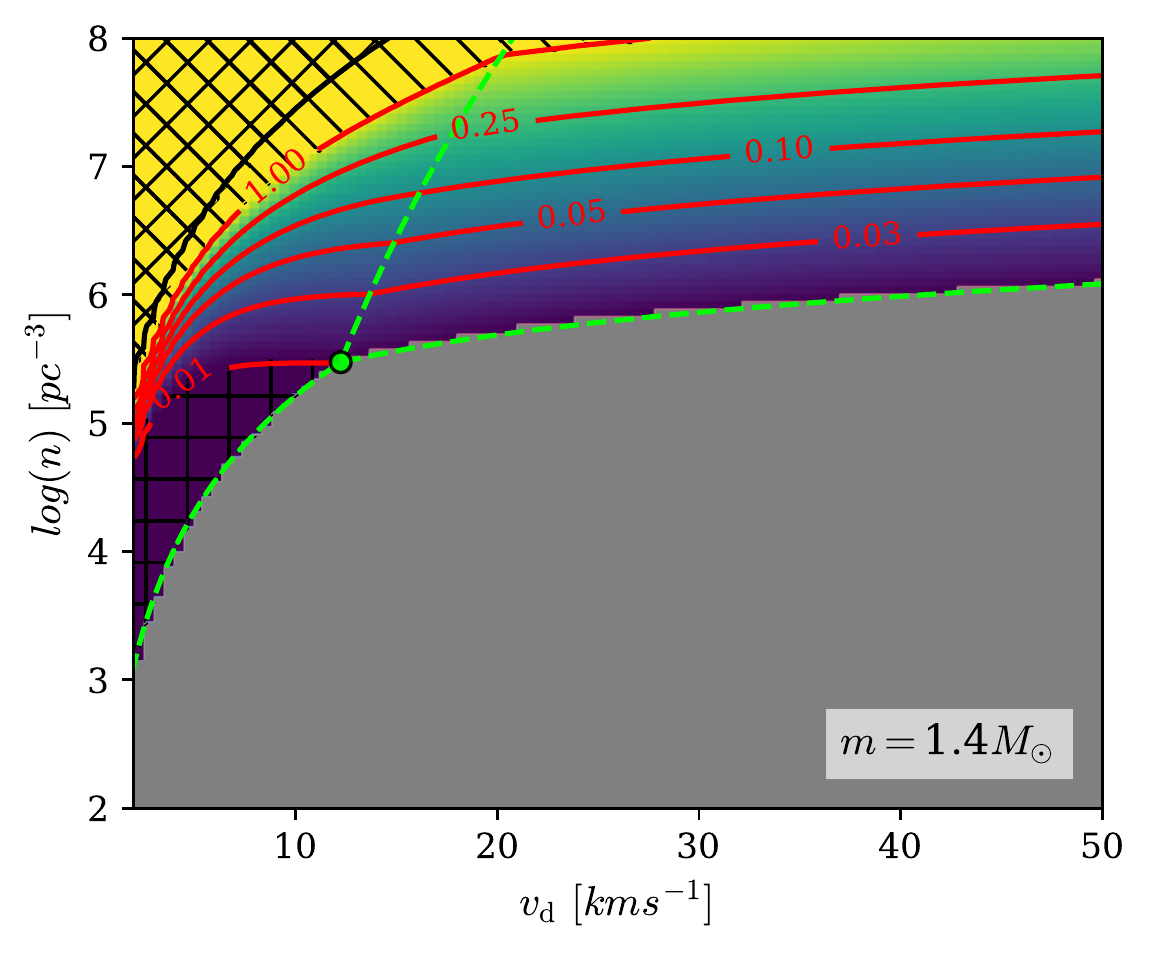}
\caption{Results from our considered case $cB$ described in Sec. \ref{sec:Time-Evolving Populations},
where the number of 1G- and 2G-object as a function of time is given by Eq. \eqref{eq:N1N2_sol_ab},
and their ratio $N_2/N_1$ by Eq. \eqref{eq:N2oN1}. We here consider solutions to $t = t_H$ for a model described by
$f_b = 0.01$, $R_{M}^{11} = 1$, $N_{t}^{ej} = 6$, and $N_{s}^{ej} = 4$, where the {\it upper} and {\it lower} plots correspond to $m=30M_{\odot}$
and $m=1.4M_{\odot}$, respectively.
The area covered by the {\it red contours} is where $0.01< N_2/N_1 < 1.0$, i.e. it is the region
that gives rise to both consistent ($<1.0$) and interesting ($>0.01$) outcomes for growing a 2G-population. In the {\it yellow `${\backslash}{\backslash}$'-hatched
area}, our formalism evaluated at $t = t_H$ breaks down as $N_2$ is here $>N_1$, where in the {\it grey area} our $N_c$ averaging approach breaks down as $N_c$ is here $< 1$.
In the {\it yellow `$X$'-hatched area} $N_1/N_1(0) < 10^{-4}$; therefore, if a system is located within this area it will `evaporate'
within a Hubble time if its initial number of BHs is $ \lesssim 10^{4}$. The {\it `$+$'-hatched area} is where $N_2/N_1 < 0.01$ and $N_c \gg 1$, and
highlights therefore systems that clearly undergo several ICs, but still end up with a relative small 2G-population.
The green separation lines are describe in Sec. \ref{sec: Outcome Regions}. Results related to this figure are described in Sec. \ref{sec:Upper Limits on 2G Objects}.}
\label{fig:N2oN1}
\end{figure}

\begin{figure}
\centering
\includegraphics[width=\columnwidth]{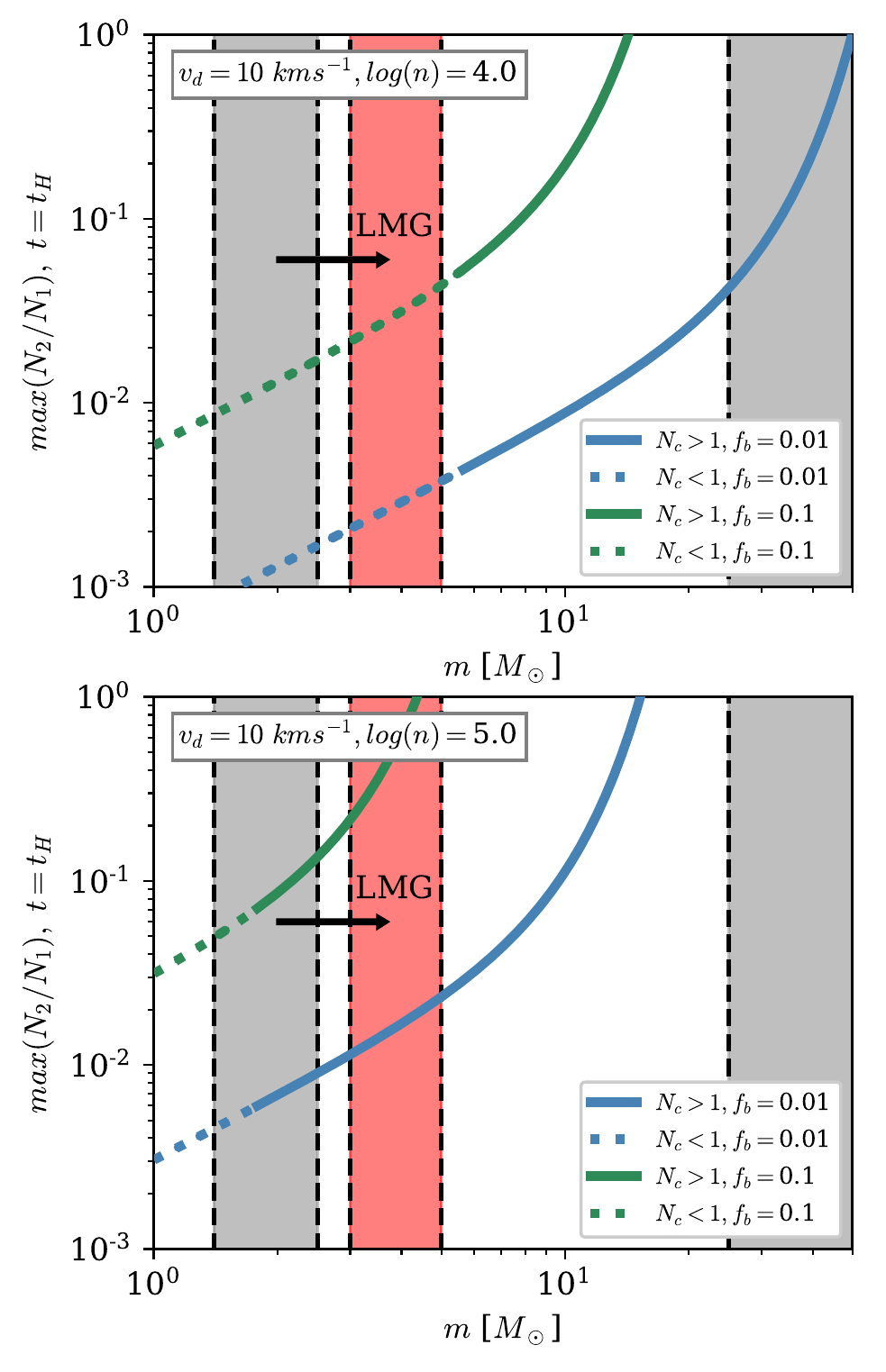}
\caption{Number of 2G-objects ($N_2$) relative to 1G-objects ($N_1$) derived for our case $cB$ using Eq. \eqref{eq:N2oN1} at $t=t_H$, as a function of $m$
for fixed $v_d$, but varying $n$ and $f_b$, as further indicated in the legends.
The two plots differ by the value of $n$, where $n = 10^{4} pc^{-3}$ and $n = 10^{5} pc^{-3}$ in the {\it upper} and {\it lower} plots, respectively.
The {\it grey bands} show the mass range for which a merger that produces a remnant with mass $\sim 2m$ will land in the corresponding mass gap, where
the {\it red band} shows the LMG (the UMG is not shown).
For example, a merger between two COs (NSs) in the lower grey band will form a merger product that lands
in the red LMG band, as further illustrated by the {\it black arrow}.
The {\it dotted lines} highlight the part of the curves for which $N_c < 1$ (2G formation is highly ineffective), where the solid lines correspond to $N_c > 1$ (2G formation is possible).}
\label{fig:N2oN1marr}
\end{figure}

Fig. \ref{fig:N2oN1} shows results related to the ratio $N_2/N_1$ given by Eq. \eqref{eq:N2oN1} evaluated at $t = t_H$,
as further described in the figure caption. As described in Sec. \ref{sec:Time-Evolving Populations}, this case represents in our model an upper limit on $N_2/N_1$.
Considering first the upper plot showing the $m=30M_{\odot}$ case, we see that for a GC with $v_d \sim 10 kms^{-1}$ a population of 1G-objects can
over a Hubble time turn into a population with $N_2/N_1 > 0.1$ if $n \gtrsim 10^{4} pc^{-3}$. Although this is an upper limit, it greatly illustrates that the length, mass, and times scales
associated with a typical cluster hosting BHs of mass $\sim 30M_{\odot}$ in the core is able to populate the upper mass gap through successive mergers of its 1G-population.
Considering now the lower plot showing results for the $1.4M_{\odot}$ case, we see that for $v_d \sim 10 kms^{-1}$ the density has to be $\gtrsim 10^{5} pc^{-3}$
to even grow a 2G-population with $N_2/N_1 > 0.01$, and $\gtrsim 10^{6} pc^{-3}$ for $N_2/N_1 > 0.1$.
From this we conclude that populating the lower mass gap through successive mergers of NSs in any reasonable astrophysical cluster seems
almost impossible, not even when we assume that the entire population is consisting of only NSs. This last assumption is in fact also
highly optimistic, as NSs will not segregate and form their own sub-cluster in the same way as BHs because their characteristic $1.4M_{\odot}$ mass is very
close to that of the ordinary stars in the cluster. As a result, NSs will exchange and interact frequently with the stellar population, which
introduces `impurities' in the IC illustrated in Fig. \ref{fig:ill_int}. The probability that two NSs merge inside the cluster is therefore
significantly smaller than what we have assumed in our considered $cB$ scenario. In comparison, the BHs have such a large mass compared to the
remaining stellar population, that they easily form their own sub-system \citep[e.g.][]{2018MNRAS.478.1844A}. In Fig. \ref{fig:N2oN1marr} we show how
these results depend more broadly on the mass $m$, where we show $N_2/N_1$ from case $cB$, as a function of $m$ for $n = 10^{4} pc^{-3}$ (top plot)
and $n = 10^{5} pc^{-3}$ (bottom plot), and two different binary fractions,
as further described in the figure caption.

Finally, we note that the real `bottle neck' in populating the lower mass gap is not directly related to the probability $P_M$ per IC
for a NS population to undergo NS-NS mergers inside their cluster. Instead, it is the time it takes for a NS-NS binary to
undergo one IC, $\tau_m$, that simply is too long for a standard cluster. This is clear from Fig. \ref{fig:N2oN1}, as the grey area, where $N_c \lesssim 1$, sets the
lower limit at $n=10^{5} pc^{-3}$ for $10kms^{-1}$. In the limit where $a_m = a_{ej}$ the number of ICs evaluated at $t_H$, $N_c(t_H) = t_H/\tau_m(a_{ej})$,
is given by,
\begin{align}
N_c(t_H)        & \approx t_{H} \left[ \frac{\pi G^2 {\Delta}^2}{\delta f_{ed}^{2}} \right] \times \frac{nm^2}{v_d^{3}} \\
                & \approx 0.8 \left(\frac{n}{10^{5}pc^{-3}}\right) \left(\frac{m}{1.4M_{\odot}}\right)^{2} \left(\frac{v_d}{10\ kms^{-1}}\right)^{-3},
\label{eq:Ncth}
\end{align}
and is indeed just around unity for NS-NS binaries for our chosen normalizations.
It is furthermore seen that $N_c(t_H)$ rapidly decreases with mass $m$ as $\propto m^{2}$. However, as seen on Fig. \ref{fig:Nmbin},
if the system is in the area for which $N_c(t_H) > 1$, the dependence on $m$ on how many in-cluster mergers a given
binary can produce within a Hubble time, $N'_{M}(t_H)$, is less sensitive to $m$, as $N_c(t_H) \times P_{m} \propto m^{8/7}$.
All in all, the limit for which $N_c(t_H) = 1$ plays therefore a crucial role for determining what systems that are able to produce a
significant 2G-population. We conclude our study below.

\section{Conclusions}
\label{sec:Conclusions}
 
We have in this paper studied the formation of 2G objects formed through 1G-1G in-cluster mergers
in dense clusters. We have in particular explored the possibility for populating the LMG ($3-5M_{\odot}$) and the UMG ($\gtrsim 45M_{\odot}$)
through the merger of BNSs and BBHs, respectively. Understanding what cluster systems that are able to populate these two
mass-gaps has wide implications for both GW astrophysics and stellar physics. For example, if nature is proven not to be able to create mass-gap BHs through
normal stellar evolution, then current and future measures of the BH mass spectrum, through e.g. GW observations, will give us insight into the formation mechanisms
of BBH mergers in clusters. On the other hand, if observations hint that stellar clusters do not contribute significantly to the observed GW merger rate,
e.g. through independent measures of the fraction of eccentric BBH mergers \citep[e.g.][]{2018PhRvD..97j3014S}, then an observed population of mass-gap objects
will hint that our single stellar models need to be revised. For these reasons, several new studies have discussed the possibility for dynamically populating these mass
gaps \citep[e.g.][]{2016ApJ...824L..12O, 2017ApJ...840L..24F, 2017PhRvD..95l4046G, 2019PhRvL.123r1101Y, 2019MNRAS.486.5008A, 2019PhRvD.100d1301G, 2019MNRAS.482...30S, 2019PhRvD.100d3027R, 2020arXiv200504243G, 2020ApJ...888L...3S, 2020ApJ...890L..20G, 2020arXiv200500023K, 2020ApJ...893...35D, 2020arXiv200400650B}.

Through a fully analytical approach we have here studied how efficient a cluster, described by a constant $v_d, n$,
can turn its initial population of $N_1$ 1G-objects into a sizable population of $N_2$ 2G-objects through in-cluster GW mergers.
We have in particular explored the upper limit on the ratio $N_2/N_1$ evaluated after a Hubble time,
as a function of $v_d, n$ and $m$ (Sec. \ref{sec:Upper Limits on 2G Objects}). Our limit is based entirely on dynamics,
and complements therefore greatly the recent study by \cite{2019PhRvD.100d1301G}, where the limit was derived from considering the magnitude of GW kicks.
From our analysis we have reached the following conclusions:

Populating the LMG through in-cluster mergers of BNSs is a very slow process for any astrophysical cluster. For example, as shown in Fig. \ref{fig:N2oN1}, even in the
highly idealized case of a GC core populated entirely by NSs, the number density $n$ has to be $>10^6\ pc^{-3}$ to reach $N_2/N_1 \sim 0.1$.
As discussed in Sec. \ref{sec:Upper Limits on 2G Objects}, not only is this density much higher than what is found for real clusters,
but NSs are also likely to mix with other stars due to their similar mass, which reduces their in-cluster merger probability further.
In fact, our results show that what really limits a NS rich core to undergo enough in-cluster mergers to populate the LMG is actually
the timescale for interactions, and not how the NSs exactly merge inside their cluster. This is seen in Fig. \ref{fig:N2oN1},
where for a NS dominated core (bottom plot) a density of $n \gtrsim 10^{5}\ pc^{-3}$ for $v_d \sim  10\ kms^{-1}$ is required to move above the grey area, i.e. for a BNS
to undergo at least 1 IC. In our described `standard picture' of dynamically assembled in-cluster mergers (Sec. \ref{sec:Cluster Model}), an efficient production of LMG objects
is therefore highly unlikely. If clusters for some reason are still observed to effectively produce LMG objects through dynamics, then more `exotic' dynamical pathways have to be evoked.
Alternatively, it could be that some clusters start out with a high BNS fraction (see Fig. \ref{fig:N2oN1marr}) that would
lead to a relative high number of 2G objects after a Hubble time. However, in that case, there would still be problems related to how fast this 2G population
can be dynamically paired up with other COs to undergo, say, observable GW mergers. Therefore, observing GW sources with at least one LMG object
formed in a cluster near the grey area in Fig. \ref{fig:N2oN1} (bottom) seems therefore highly unlikely.
 
Populating the UMG is in comparison much easier, e.g., in Fig. \ref{fig:N2oN1} (top) it is clearly seen that reaching values of $N_2/N_1 \sim 0.1$ only requires clusters with a central
density of $\sim 10^{4}\ pc^{-3}$. This is a much more reasonable magnitude, which leads us to conclude that populating the UMG in clusters is relatively easy, at least dynamically,
without introducing any non-standard pathways. Our model even implies that for $n \gtrsim 10^{4}\ pc^{-3}$ there is high
probability for the initial 1G population to turn almost entirely into a 2G population. Our models are not able to accurately describe this scenario, but it does at least hint that in
moderate dense clusters in-cluster mergers can be highly effective in changing the initial mass function. This has great implications for 3. generation GW observations where
will see every BBH merger within our observable patch as a function of redshift.

Finally, we note that a few studies that were completed while our present study was underway point towards similar conclusions to what we
have her. For example, in \cite{2020ApJ...888L..10Y} it was shown using a fully numerical approach that the rate of BNS mergers originating from GCs is low,
where both \cite{2019PhRvD.100d3027R} and \cite{2020arXiv200400650B} illustrated that populating the UMG definitely seems possible.
However, other studies still keep the question open to what degree the LMG can be populated in clusters \citep[e.g.][]{2020PhRvD.101j3036G}. The topic is therefore highly
rich and interesting, and our study greatly compliments this recent literature with the first set of closed form solutions that encapsulate all the correct scalings
and relations of the problem. We note here that standard brute-force $N$-body techniques are still too slow at evolving high density clusters,
which is why we and others explore how to solve this problem using approximate schemes (see also work by \cite{2019MNRAS.486.5008A, 2020MNRAS.492.2936A}).
We are currently working on a self consistent hybrid scheme that will enable us to correctly evolve a full mass distribution.
Our present paper plays a crucial role in providing the first steps in this highly relevant and timely topic.

\acknowledgments
The authors thank the Yukawa Institute for Theoretical Physics at Kyoto University,
and the organizers of the workshop YKIS2019 `Black Holes and Neutron Stars with Gravitational Waves',
where many useful conversations took place. It is also a pleasure to thank Kyle Kremer for enlightening discussions.
JS acknowledges support from the Lyman Spitzer Fellowship and the European Unions Horizon 2020
research and innovation programme under the Marie Sklodowska-Curie grant agreement No. 844629.
KH acknowledges support from the Lyman Spitzer Fellowship.

\bibliographystyle{aasjournal}
\bibliography{NbodyTides_papers}

\begin{thebibliography}{}
\expandafter\ifx\csname natexlab\endcsname\relax\def\natexlab#1{#1}\fi
\providecommand{\url}[1]{\href{#1}{#1}}
\providecommand{\dodoi}[1]{doi:~\href{http://doi.org/#1}{\nolinkurl{#1}}}
\providecommand{\doeprint}[1]{\href{http://ascl.net/#1}{\nolinkurl{http://ascl.net/#1}}}
\providecommand{\doarXiv}[1]{\href{https://arxiv.org/abs/#1}{\nolinkurl{https://arxiv.org/abs/#1}}}

\bibitem[{{Aarseth} \& {Heggie}(1976)}]{1976A&A....53..259A}
{Aarseth}, S.~J., \& {Heggie}, D.~C. 1976, \aap, 53, 259

\bibitem[{{Abbott} {et~al.}(2016{\natexlab{a}}){Abbott}, {Abbott}, {Abbott},
  {Abernathy}, {Acernese}, {Ackley}, {Adams}, {Adams}, {Addesso}, {Adhikari},
  \& et~al.}]{2016PhRvL.116f1102A}
{Abbott}, B.~P., {Abbott}, R., {Abbott}, T.~D., {et~al.} 2016{\natexlab{a}},
  Physical Review Letters, 116, 061102, \dodoi{10.1103/PhysRevLett.116.061102}

\bibitem[{{Abbott} {et~al.}(2016{\natexlab{b}}){Abbott}, {Abbott}, {Abbott},
  {Abernathy}, {Acernese}, {Ackley}, {Adams}, {Adams}, {Addesso}, {Adhikari},
  \& et~al.}]{2016PhRvL.116x1103A}
---. 2016{\natexlab{b}}, Physical Review Letters, 116, 241103,
  \dodoi{10.1103/PhysRevLett.116.241103}

\bibitem[{{Abbott} {et~al.}(2016{\natexlab{c}}){Abbott}, {Abbott}, {Abbott},
  {Abernathy}, {Acernese}, {Ackley}, {Adams}, {Adams}, {Addesso}, {Adhikari},
  \& et~al.}]{2016PhRvX...6d1015A}
---. 2016{\natexlab{c}}, Physical Review X, 6, 041015,
  \dodoi{10.1103/PhysRevX.6.041015}

\bibitem[{{Abbott} {et~al.}(2017{\natexlab{a}}){Abbott}, {Abbott}, {Abbott},
  {Acernese}, {Ackley}, {Adams}, {Adams}, {Addesso}, {Adhikari}, {Adya}, \&
  et~al.}]{2017PhRvL.118v1101A}
---. 2017{\natexlab{a}}, Physical Review Letters, 118, 221101,
  \dodoi{10.1103/PhysRevLett.118.221101}

\bibitem[{{Abbott} {et~al.}(2017{\natexlab{b}}){Abbott}, {Abbott}, {Abbott},
  {Acernese}, {Ackley}, {Adams}, {Adams}, {Addesso}, {Adhikari}, {Adya}, \&
  et~al.}]{2017PhRvL.119n1101A}
---. 2017{\natexlab{b}}, Physical Review Letters, 119, 141101,
  \dodoi{10.1103/PhysRevLett.119.141101}

\bibitem[{{Abbott} {et~al.}(2017{\natexlab{c}}){Abbott}, {Abbott}, {Abbott},
  {Acernese}, {Ackley}, {Adams}, {Adams}, {Addesso}, {Adhikari}, {Adya}, \&
  et~al.}]{2017PhRvL.119p1101A}
---. 2017{\natexlab{c}}, Physical Review Letters, 119, 161101,
  \dodoi{10.1103/PhysRevLett.119.161101}

\bibitem[{{Andrews} \& {Mandel}(2019)}]{2019ApJ...880L...8A}
{Andrews}, J.~J., \& {Mandel}, I. 2019, \apjl, 880, L8,
  \dodoi{10.3847/2041-8213/ab2ed1}

\bibitem[{{Antonini} {et~al.}(2016){Antonini}, {Chatterjee}, {Rodriguez},
  {Morscher}, {Pattabiraman}, {Kalogera}, \& {Rasio}}]{2016ApJ...816...65A}
{Antonini}, F., {Chatterjee}, S., {Rodriguez}, C.~L., {et~al.} 2016, \apj, 816,
  65, \dodoi{10.3847/0004-637X/816/2/65}

\bibitem[{{Antonini} \& {Gieles}(2020)}]{2020MNRAS.492.2936A}
{Antonini}, F., \& {Gieles}, M. 2020, \mnras, 492, 2936,
  \dodoi{10.1093/mnras/stz3584}

\bibitem[{{Antonini} {et~al.}(2019){Antonini}, {Gieles}, \&
  {Gualandris}}]{2019MNRAS.486.5008A}
{Antonini}, F., {Gieles}, M., \& {Gualandris}, A. 2019, \mnras, 486, 5008,
  \dodoi{10.1093/mnras/stz1149}

\bibitem[{{Antonini} \& {Rasio}(2016)}]{2016ApJ...831..187A}
{Antonini}, F., \& {Rasio}, F.~A. 2016, \apj, 831, 187,
  \dodoi{10.3847/0004-637X/831/2/187}

\bibitem[{{Antonini} {et~al.}(2018){Antonini}, {Rodriguez}, {Petrovich}, \&
  {Fischer}}]{2018MNRAS.480L..58A}
{Antonini}, F., {Rodriguez}, C.~L., {Petrovich}, C., \& {Fischer}, C.~L. 2018,
  \mnras, 480, L58, \dodoi{10.1093/mnrasl/sly126}

\bibitem[{{Antonini} {et~al.}(2017){Antonini}, {Toonen}, \&
  {Hamers}}]{2017ApJ...841...77A}
{Antonini}, F., {Toonen}, S., \& {Hamers}, A.~S. 2017, \apj, 841, 77,
  \dodoi{10.3847/1538-4357/aa6f5e}

\bibitem[{{Askar} {et~al.}(2018){Askar}, {Arca Sedda}, \&
  {Giersz}}]{2018MNRAS.478.1844A}
{Askar}, A., {Arca Sedda}, M., \& {Giersz}, M. 2018, \mnras, 478, 1844,
  \dodoi{10.1093/mnras/sty1186}

\bibitem[{{Askar} {et~al.}(2017){Askar}, {Szkudlarek}, {Gondek-Rosi{\'n}ska},
  {Giersz}, \& {Bulik}}]{2017MNRAS.464L..36A}
{Askar}, A., {Szkudlarek}, M., {Gondek-Rosi{\'n}ska}, D., {Giersz}, M., \&
  {Bulik}, T. 2017, \mnras, 464, L36, \dodoi{10.1093/mnrasl/slw177}

\bibitem[{{Bae} {et~al.}(2014){Bae}, {Kim}, \& {Lee}}]{2014MNRAS.440.2714B}
{Bae}, Y.-B., {Kim}, C., \& {Lee}, H.~M. 2014, \mnras, 440, 2714,
  \dodoi{10.1093/mnras/stu381}

\bibitem[{{Baibhav} {et~al.}(2020){Baibhav}, {Gerosa}, {Berti}, {Wong},
  {Helfer}, \& {Mould}}]{2020arXiv200400650B}
{Baibhav}, V., {Gerosa}, D., {Berti}, E., {et~al.} 2020, arXiv e-prints,
  arXiv:2004.00650.
\newblock \doarXiv{2004.00650}

\bibitem[{{Bailyn} {et~al.}(1998){Bailyn}, {Jain}, {Coppi}, \&
  {Orosz}}]{Bailyn1998ApJ}
{Bailyn}, C.~D., {Jain}, R.~K., {Coppi}, P., \& {Orosz}, J.~A. 1998, \apj, 499,
  367, \dodoi{10.1086/305614}

\bibitem[{{Banerjee} {et~al.}(2010){Banerjee}, {Baumgardt}, \&
  {Kroupa}}]{2010MNRAS.402..371B}
{Banerjee}, S., {Baumgardt}, H., \& {Kroupa}, P. 2010, \mnras, 402, 371,
  \dodoi{10.1111/j.1365-2966.2009.15880.x}

\bibitem[{{Bartos} {et~al.}(2017){Bartos}, {Kocsis}, {Haiman}, \&
  {M{\'a}rka}}]{2017ApJ...835..165B}
{Bartos}, I., {Kocsis}, B., {Haiman}, Z., \& {M{\'a}rka}, S. 2017, \apj, 835,
  165, \dodoi{10.3847/1538-4357/835/2/165}

\bibitem[{{Belczynski} {et~al.}(2016{\natexlab{a}}){Belczynski}, {Holz},
  {Bulik}, \& {O'Shaughnessy}}]{2016Natur.534..512B}
{Belczynski}, K., {Holz}, D.~E., {Bulik}, T., \& {O'Shaughnessy}, R.
  2016{\natexlab{a}}, \nat, 534, 512, \dodoi{10.1038/nature18322}

\bibitem[{{Belczynski} {et~al.}(2016{\natexlab{b}}){Belczynski}, {Repetto},
  {Holz}, {O'Shaughnessy}, {Bulik}, {Berti}, {Fryer}, \&
  {Dominik}}]{2016ApJ...819..108B}
{Belczynski}, K., {Repetto}, S., {Holz}, D.~E., {et~al.} 2016{\natexlab{b}},
  \apj, 819, 108, \dodoi{10.3847/0004-637X/819/2/108}

\bibitem[{{Berti} {et~al.}(2007){Berti}, {Cardoso}, {Gonzalez}, {Sperhake},
  {Hannam}, {Husa}, \& {Br{\"u}gmann}}]{2007PhRvD..76f4034B}
{Berti}, E., {Cardoso}, V., {Gonzalez}, J.~A., {et~al.} 2007, \prd, 76, 064034,
  \dodoi{10.1103/PhysRevD.76.064034}

\bibitem[{{Bird} {et~al.}(2016){Bird}, {Cholis}, {Mu{\~n}oz},
  {Ali-Ha{\"i}moud}, {Kamionkowski}, {Kovetz}, {Raccanelli}, \&
  {Riess}}]{2016PhRvL.116t1301B}
{Bird}, S., {Cholis}, I., {Mu{\~n}oz}, J.~B., {et~al.} 2016, Physical Review
  Letters, 116, 201301, \dodoi{10.1103/PhysRevLett.116.201301}

\bibitem[{{Carr} {et~al.}(2016){Carr}, {K{\"u}hnel}, \&
  {Sandstad}}]{2016PhRvD..94h3504C}
{Carr}, B., {K{\"u}hnel}, F., \& {Sandstad}, M. 2016, \prd, 94, 083504,
  \dodoi{10.1103/PhysRevD.94.083504}

\bibitem[{{Chen} \& {Amaro-Seoane}(2017)}]{2017ApJ...842L...2C}
{Chen}, X., \& {Amaro-Seoane}, P. 2017, \apjl, 842, L2,
  \dodoi{10.3847/2041-8213/aa74ce}

\bibitem[{{Cholis} {et~al.}(2016){Cholis}, {Kovetz}, {Ali-Ha{\"i}moud}, {Bird},
  {Kamionkowski}, {Mu{\~n}oz}, \& {Raccanelli}}]{2016PhRvD..94h4013C}
{Cholis}, I., {Kovetz}, E.~D., {Ali-Ha{\"i}moud}, Y., {et~al.} 2016, \prd, 94,
  084013, \dodoi{10.1103/PhysRevD.94.084013}

\bibitem[{{Doctor} {et~al.}(2020){Doctor}, {Wysocki}, {O'Shaughnessy}, {Holz},
  \& {Farr}}]{2020ApJ...893...35D}
{Doctor}, Z., {Wysocki}, D., {O'Shaughnessy}, R., {Holz}, D.~E., \& {Farr}, B.
  2020, \apj, 893, 35, \dodoi{10.3847/1538-4357/ab7fac}

\bibitem[{{Dominik} {et~al.}(2012){Dominik}, {Belczynski}, {Fryer}, {Holz},
  {Berti}, {Bulik}, {Mandel}, \& {O'Shaughnessy}}]{2012ApJ...759...52D}
{Dominik}, M., {Belczynski}, K., {Fryer}, C., {et~al.} 2012, \apj, 759, 52,
  \dodoi{10.1088/0004-637X/759/1/52}

\bibitem[{{Dominik} {et~al.}(2013){Dominik}, {Belczynski}, {Fryer}, {Holz},
  {Berti}, {Bulik}, {Mandel}, \& {O'Shaughnessy}}]{2013ApJ...779...72D}
---. 2013, \apj, 779, 72, \dodoi{10.1088/0004-637X/779/1/72}

\bibitem[{{Dominik} {et~al.}(2015){Dominik}, {Berti}, {O'Shaughnessy},
  {Mandel}, {Belczynski}, {Fryer}, {Holz}, {Bulik}, \&
  {Pannarale}}]{2015ApJ...806..263D}
{Dominik}, M., {Berti}, E., {O'Shaughnessy}, R., {et~al.} 2015, \apj, 806, 263,
  \dodoi{10.1088/0004-637X/806/2/263}

\bibitem[{{D'Orazio} \& {Loeb}(2017)}]{DOrazioLoeb:2017}
{D'Orazio}, D.~J., \& {Loeb}, A. 2017, ArXiv e-prints.
\newblock \doarXiv{1706.04211}

\bibitem[{{Farmer} {et~al.}(2019){Farmer}, {Renzo}, {de Mink}, {Marchant}, \&
  {Justham}}]{Farmer2019ApJ}
{Farmer}, R., {Renzo}, M., {de Mink}, S.~E., {Marchant}, P., \& {Justham}, S.
  2019, \apj, 887, 53, \dodoi{10.3847/1538-4357/ab518b}

\bibitem[{{Farr} {et~al.}(2019){Farr}, {Fishbach}, {Ye}, \&
  {Holz}}]{2019ApJ...883L..42F}
{Farr}, W.~M., {Fishbach}, M., {Ye}, J., \& {Holz}, D.~E. 2019, \apjl, 883,
  L42, \dodoi{10.3847/2041-8213/ab4284}

\bibitem[{{Farr} {et~al.}(2011){Farr}, {Sravan}, {Cantrell}, {Kreidberg},
  {Bailyn}, {Mandel}, \& {Kalogera}}]{farr2011ApJ}
{Farr}, W.~M., {Sravan}, N., {Cantrell}, A., {et~al.} 2011, \apj, 741, 103,
  \dodoi{10.1088/0004-637X/741/2/103}

\bibitem[{{Fishbach} {et~al.}(2017){Fishbach}, {Holz}, \&
  {Farr}}]{2017ApJ...840L..24F}
{Fishbach}, M., {Holz}, D.~E., \& {Farr}, B. 2017, \apjl, 840, L24,
  \dodoi{10.3847/2041-8213/aa7045}

\bibitem[{{Fragione} \& {Bromberg}(2019)}]{2019arXiv190309659F}
{Fragione}, G., \& {Bromberg}, O. 2019, arXiv e-prints, arXiv:1903.09659.
\newblock \doarXiv{1903.09659}

\bibitem[{{Fragione} \& {Kocsis}(2018)}]{2018PhRvL.121p1103F}
{Fragione}, G., \& {Kocsis}, B. 2018, \prl, 121, 161103,
  \dodoi{10.1103/PhysRevLett.121.161103}

\bibitem[{{Fragione} \& {Kocsis}(2019)}]{2019MNRAS.486.4781F}
---. 2019, \mnras, 486, 4781, \dodoi{10.1093/mnras/stz1175}

\bibitem[{{Fragione} \& {Kocsis}(2020)}]{2020MNRAS.493.3920F}
---. 2020, \mnras, 493, 3920, \dodoi{10.1093/mnras/staa443}

\bibitem[{{Fragione} \& {Loeb}(2019)}]{2019MNRAS.486.4443F}
{Fragione}, G., \& {Loeb}, A. 2019, \mnras, 486, 4443,
  \dodoi{10.1093/mnras/stz1131}

\bibitem[{{Gayathri} {et~al.}(2020){Gayathri}, {Bartos}, {Haiman}, {Klimenko},
  {Kocsis}, {M{\'a}rka}, \& {Yang}}]{2020ApJ...890L..20G}
{Gayathri}, V., {Bartos}, I., {Haiman}, Z., {et~al.} 2020, \apjl, 890, L20,
  \dodoi{10.3847/2041-8213/ab745d}

\bibitem[{{Gerosa} \& {Berti}(2017)}]{2017PhRvD..95l4046G}
{Gerosa}, D., \& {Berti}, E. 2017, \prd, 95, 124046,
  \dodoi{10.1103/PhysRevD.95.124046}

\bibitem[{{Gerosa} \& {Berti}(2019)}]{2019PhRvD.100d1301G}
---. 2019, \prd, 100, 041301, \dodoi{10.1103/PhysRevD.100.041301}

\bibitem[{{Gerosa} {et~al.}(2020){Gerosa}, {Vitale}, \&
  {Berti}}]{2020arXiv200504243G}
{Gerosa}, D., {Vitale}, S., \& {Berti}, E. 2020, arXiv e-prints,
  arXiv:2005.04243.
\newblock \doarXiv{2005.04243}

\bibitem[{{Giersz} {et~al.}(2015){Giersz}, {Leigh}, {Hypki}, {L{\"u}tzgendorf},
  \& {Askar}}]{2015MNRAS.454.3150G}
{Giersz}, M., {Leigh}, N., {Hypki}, A., {L{\"u}tzgendorf}, N., \& {Askar}, A.
  2015, \mnras, 454, 3150, \dodoi{10.1093/mnras/stv2162}

\bibitem[{G{\"u}ltekin {et~al.}(2004)G{\"u}ltekin, Miller, \&
  Hamilton}]{2004ApJ...616..221G}
G{\"u}ltekin, K., Miller, M.~C., \& Hamilton, D.~P. 2004, \apj, 616, 221

\bibitem[{G{\"u}ltekin {et~al.}(2006)G{\"u}ltekin, Miller, \&
  Hamilton}]{2006ApJ...640..156G}
---. 2006, \apj, 640, 156

\bibitem[{{Gupta} {et~al.}(2020){Gupta}, {Gerosa}, {Arun}, {Berti}, {Farr}, \&
  {Sathyaprakash}}]{2020PhRvD.101j3036G}
{Gupta}, A., {Gerosa}, D., {Arun}, K.~G., {et~al.} 2020, \prd, 101, 103036,
  \dodoi{10.1103/PhysRevD.101.103036}

\bibitem[{{Hamers} {et~al.}(2018){Hamers}, {Bar-Or}, {Petrovich}, \&
  {Antonini}}]{2018ApJ...865....2H}
{Hamers}, A.~S., {Bar-Or}, B., {Petrovich}, C., \& {Antonini}, F. 2018, \apj,
  865, 2, \dodoi{10.3847/1538-4357/aadae2}

\bibitem[{{Hamers} \& {Samsing}(2019{\natexlab{a}})}]{2019MNRAS.487.5630H}
{Hamers}, A.~S., \& {Samsing}, J. 2019{\natexlab{a}}, \mnras, 487, 5630,
  \dodoi{10.1093/mnras/stz1646}

\bibitem[{{Hamers} \& {Samsing}(2019{\natexlab{b}})}]{2019MNRAS.488.5192H}
---. 2019{\natexlab{b}}, \mnras, 488, 5192, \dodoi{10.1093/mnras/stz2029}

\bibitem[{{Hamers} \& {Samsing}(2020)}]{2020MNRAS.494..850H}
---. 2020, \mnras, 494, 850, \dodoi{10.1093/mnras/staa691}

\bibitem[{{Hamers} \& {Thompson}(2019)}]{2019ApJ...883...23H}
{Hamers}, A.~S., \& {Thompson}, T.~A. 2019, \apj, 883, 23,
  \dodoi{10.3847/1538-4357/ab3b06}

\bibitem[{Heggie(1975)}]{Heggie:1975uy}
Heggie, D.~C. 1975, \mnras, 173, 729

\bibitem[{{H{\'e}nault-Brunet} {et~al.}(2020){H{\'e}nault-Brunet}, {Gieles},
  {Strader}, {Peuten}, {Balbinot}, \& {Douglas}}]{2020MNRAS.491..113H}
{H{\'e}nault-Brunet}, V., {Gieles}, M., {Strader}, J., {et~al.} 2020, \mnras,
  491, 113, \dodoi{10.1093/mnras/stz2995}

\bibitem[{{Hoang} {et~al.}(2017){Hoang}, {Naoz}, {Kocsis}, {Rasio}, \&
  {Dosopoulou}}]{2017arXiv170609896H}
{Hoang}, B.-M., {Naoz}, S., {Kocsis}, B., {Rasio}, F.~A., \& {Dosopoulou}, F.
  2017, ArXiv e-prints.
\newblock \doarXiv{1706.09896}

\bibitem[{{Hong} \& {Lee}(2015)}]{2015MNRAS.448..754H}
{Hong}, J., \& {Lee}, H.~M. 2015, \mnras, 448, 754,
  \dodoi{10.1093/mnras/stv035}

\bibitem[{{Hotokezaka} \& {Piran}(2017)}]{Hotokezaka2017ApJ}
{Hotokezaka}, K., \& {Piran}, T. 2017, \apj, 842, 111,
  \dodoi{10.3847/1538-4357/aa6f61}

\bibitem[{Hut \& Bahcall(1983)}]{Hut:1983js}
Hut, P., \& Bahcall, J.~N. 1983, \apj, 268, 319

\bibitem[{{Janiuk} {et~al.}(2017){Janiuk}, {Bejger}, {Charzy{\'n}ski}, \&
  {Sukova}}]{Janiuk+2017}
{Janiuk}, A., {Bejger}, M., {Charzy{\'n}ski}, S., \& {Sukova}, P. 2017, ArXiv
  e-prints, 51, 7, \dodoi{10.1016/j.newast.2016.08.002}

\bibitem[{{Kalogera}(2000)}]{2000ApJ...541..319K}
{Kalogera}, V. 2000, \apj, 541, 319, \dodoi{10.1086/309400}

\bibitem[{{Kimball} {et~al.}(2020){Kimball}, {Talbot}, {Berry}, {Carney},
  {Zevin}, {Thrane}, \& {Kalogera}}]{2020arXiv200500023K}
{Kimball}, C., {Talbot}, C., {Berry}, C. P.~L., {et~al.} 2020, arXiv e-prints,
  arXiv:2005.00023.
\newblock \doarXiv{2005.00023}

\bibitem[{{Kinugawa} {et~al.}(2014){Kinugawa}, {Inayoshi}, {Hotokezaka},
  {Nakauchi}, \& {Nakamura}}]{Kinugawa2014MNRAS}
{Kinugawa}, T., {Inayoshi}, K., {Hotokezaka}, K., {Nakauchi}, D., \&
  {Nakamura}, T. 2014, \mnras, 442, 2963, \dodoi{10.1093/mnras/stu1022}

\bibitem[{{K{\i}z{\i}ltan} {et~al.}(2017){K{\i}z{\i}ltan}, {Baumgardt}, \&
  {Loeb}}]{2017Natur.542..203K}
{K{\i}z{\i}ltan}, B., {Baumgardt}, H., \& {Loeb}, A. 2017, \nat, 542, 203,
  \dodoi{10.1038/nature21361}

\bibitem[{{Kremer} {et~al.}(2019{\natexlab{a}}){Kremer}, {Lu}, {Rodriguez},
  {Lachat}, \& {Rasio}}]{2019ApJ...881...75K}
{Kremer}, K., {Lu}, W., {Rodriguez}, C.~L., {Lachat}, M., \& {Rasio}, F.~A.
  2019{\natexlab{a}}, \apj, 881, 75, \dodoi{10.3847/1538-4357/ab2e0c}

\bibitem[{{Kremer} {et~al.}(2020){Kremer}, {Ye}, {Chatterjee}, {Rodriguez}, \&
  {Rasio}}]{2020IAUS..351..357K}
{Kremer}, K., {Ye}, C.~S., {Chatterjee}, S., {Rodriguez}, C.~L., \& {Rasio},
  F.~A. 2020, in IAU Symposium, Vol. 351, IAU Symposium, ed. A.~{Bragaglia},
  M.~{Davies}, A.~{Sills}, \& E.~{Vesperini}, 357--366,
  \dodoi{10.1017/S1743921319007269}

\bibitem[{{Kremer} {et~al.}(2019{\natexlab{b}}){Kremer}, {Rodriguez},
  {Amaro-Seoane}, {Breivik}, {Chatterjee}, {Katz}, {Larson}, {Rasio},
  {Samsing}, {Ye}, \& {Zevin}}]{2019PhRvD..99f3003K}
{Kremer}, K., {Rodriguez}, C.~L., {Amaro-Seoane}, P., {et~al.}
  2019{\natexlab{b}}, \prd, 99, 063003, \dodoi{10.1103/PhysRevD.99.063003}

\bibitem[{{Leung} {et~al.}(2019){Leung}, {Nomoto}, \&
  {Blinnikov}}]{Leung2019ApJ}
{Leung}, S.-C., {Nomoto}, K., \& {Blinnikov}, S. 2019, \apj, 887, 72,
  \dodoi{10.3847/1538-4357/ab4fe5}

\bibitem[{{Liu} \& {Lai}(2017)}]{2017ApJ...846L..11L}
{Liu}, B., \& {Lai}, D. 2017, \apjl, 846, L11, \dodoi{10.3847/2041-8213/aa8727}

\bibitem[{{Liu} \& {Lai}(2018)}]{2018ApJ...863...68L}
---. 2018, \apj, 863, 68, \dodoi{10.3847/1538-4357/aad09f}

\bibitem[{{Liu} \& {Lai}(2019)}]{2019MNRAS.483.4060L}
---. 2019, \mnras, 483, 4060, \dodoi{10.1093/mnras/sty3432}

\bibitem[{{Liu} {et~al.}(2019){Liu}, {Lai}, \& {Wang}}]{2019ApJ...881...41L}
{Liu}, B., {Lai}, D., \& {Wang}, Y.-H. 2019, \apj, 881, 41,
  \dodoi{10.3847/1538-4357/ab2dfb}

\bibitem[{{Loeb}(2016)}]{Loeb:2016}
{Loeb}, A. 2016, \apjl, 819, L21, \dodoi{10.3847/2041-8205/819/2/L21}

\bibitem[{{Lopez} {et~al.}(2019){Lopez}, {Batta}, {Ramirez-Ruiz}, {Martinez},
  \& {Samsing}}]{2019ApJ...877...56L}
{Lopez}, Martin, J., {Batta}, A., {Ramirez-Ruiz}, E., {Martinez}, I., \&
  {Samsing}, J. 2019, \apj, 877, 56, \dodoi{10.3847/1538-4357/ab1842}

\bibitem[{{McKernan} {et~al.}(2017){McKernan}, {Ford}, {Bellovary}, {Leigh},
  {Haiman}, {Kocsis}, {Lyra}, {MacLow}, {Metzger}, {O'Dowd}, {Endlich}, \&
  {Rosen}}]{2017arXiv170207818M}
{McKernan}, B., {Ford}, K.~E.~S., {Bellovary}, J., {et~al.} 2017, ArXiv
  e-prints.
\newblock \doarXiv{1702.07818}

\bibitem[{{Miller} \& {Davies}(2012)}]{2012ApJ...755...81M}
{Miller}, M.~C., \& {Davies}, M.~B. 2012, \apj, 755, 81,
  \dodoi{10.1088/0004-637X/755/1/81}

\bibitem[{{Murguia-Berthier} {et~al.}(2017){Murguia-Berthier}, {MacLeod},
  {Ramirez-Ruiz}, {Antoni}, \& {Macias}}]{2017ApJ...845..173M}
{Murguia-Berthier}, A., {MacLeod}, M., {Ramirez-Ruiz}, E., {Antoni}, A., \&
  {Macias}, P. 2017, \apj, 845, 173, \dodoi{10.3847/1538-4357/aa8140}

\bibitem[{{Naoz}(2016)}]{2016ARA&A..54..441N}
{Naoz}, S. 2016, \araa, 54, 441, \dodoi{10.1146/annurev-astro-081915-023315}

\bibitem[{{Naoz} {et~al.}(2013){Naoz}, {Kocsis}, {Loeb}, \&
  {Yunes}}]{2013ApJ...773..187N}
{Naoz}, S., {Kocsis}, B., {Loeb}, A., \& {Yunes}, N. 2013, \apj, 773, 187,
  \dodoi{10.1088/0004-637X/773/2/187}

\bibitem[{{O'Leary} {et~al.}(2009){O'Leary}, {Kocsis}, \&
  {Loeb}}]{2009MNRAS.395.2127O}
{O'Leary}, R.~M., {Kocsis}, B., \& {Loeb}, A. 2009, \mnras, 395, 2127,
  \dodoi{10.1111/j.1365-2966.2009.14653.x}

\bibitem[{{O'Leary} {et~al.}(2016){O'Leary}, {Meiron}, \&
  {Kocsis}}]{2016ApJ...824L..12O}
{O'Leary}, R.~M., {Meiron}, Y., \& {Kocsis}, B. 2016, \apjl, 824, L12,
  \dodoi{10.3847/2041-8205/824/1/L12}

\bibitem[{{{\"O}zel} {et~al.}(2010){{\"O}zel}, {Psaltis}, {Narayan}, \&
  {McClintock}}]{Ozel2010ApJ}
{{\"O}zel}, F., {Psaltis}, D., {Narayan}, R., \& {McClintock}, J.~E. 2010,
  \apj, 725, 1918, \dodoi{10.1088/0004-637X/725/2/1918}

\bibitem[{{Park} {et~al.}(2017){Park}, {Kim}, {Lee}, {Bae}, \&
  {Belczynski}}]{2017MNRAS.469.4665P}
{Park}, D., {Kim}, C., {Lee}, H.~M., {Bae}, Y.-B., \& {Belczynski}, K. 2017,
  \mnras, 469, 4665, \dodoi{10.1093/mnras/stx1015}

\bibitem[{{Piran} \& {Piran}(2020)}]{Piran2020ApJ}
{Piran}, Z., \& {Piran}, T. 2020, \apj, 892, 64,
  \dodoi{10.3847/1538-4357/ab792a}

\bibitem[{Portegies~Zwart \& McMillan(2000)}]{2000ApJ...528L..17P}
Portegies~Zwart, S.~F., \& McMillan, S. L.~W. 2000, \apj, 528, L17

\bibitem[{{Randall} \& {Xianyu}(2018{\natexlab{a}})}]{2018ApJ...864..134R}
{Randall}, L., \& {Xianyu}, Z.-Z. 2018{\natexlab{a}}, \apj, 864, 134,
  \dodoi{10.3847/1538-4357/aad7fe}

\bibitem[{{Randall} \& {Xianyu}(2018{\natexlab{b}})}]{2018ApJ...853...93R}
---. 2018{\natexlab{b}}, \apj, 853, 93, \dodoi{10.3847/1538-4357/aaa1a2}

\bibitem[{{Rodriguez} {et~al.}(2018){Rodriguez}, {Amaro-Seoane}, {Chatterjee},
  {Kremer}, {Rasio}, {Samsing}, {Ye}, \& {Zevin}}]{2018PhRvD..98l3005R}
{Rodriguez}, C.~L., {Amaro-Seoane}, P., {Chatterjee}, S., {et~al.} 2018, \prd,
  98, 123005, \dodoi{10.1103/PhysRevD.98.123005}

\bibitem[{{Rodriguez} \& {Antonini}(2018)}]{2018ApJ...863....7R}
{Rodriguez}, C.~L., \& {Antonini}, F. 2018, \apj, 863, 7,
  \dodoi{10.3847/1538-4357/aacea4}

\bibitem[{{Rodriguez} {et~al.}(2016{\natexlab{a}}){Rodriguez}, {Chatterjee}, \&
  {Rasio}}]{2016PhRvD..93h4029R}
{Rodriguez}, C.~L., {Chatterjee}, S., \& {Rasio}, F.~A. 2016{\natexlab{a}},
  \prd, 93, 084029, \dodoi{10.1103/PhysRevD.93.084029}

\bibitem[{{Rodriguez} {et~al.}(2016{\natexlab{b}}){Rodriguez}, {Haster},
  {Chatterjee}, {Kalogera}, \& {Rasio}}]{2016ApJ...824L...8R}
{Rodriguez}, C.~L., {Haster}, C.-J., {Chatterjee}, S., {Kalogera}, V., \&
  {Rasio}, F.~A. 2016{\natexlab{b}}, \apjl, 824, L8,
  \dodoi{10.3847/2041-8205/824/1/L8}

\bibitem[{{Rodriguez} {et~al.}(2015){Rodriguez}, {Morscher}, {Pattabiraman},
  {Chatterjee}, {Haster}, \& {Rasio}}]{2015PhRvL.115e1101R}
{Rodriguez}, C.~L., {Morscher}, M., {Pattabiraman}, B., {et~al.} 2015, Physical
  Review Letters, 115, 051101, \dodoi{10.1103/PhysRevLett.115.051101}

\bibitem[{{Rodriguez} {et~al.}(2019){Rodriguez}, {Zevin}, {Amaro-Seoane},
  {Chatterjee}, {Kremer}, {Rasio}, \& {Ye}}]{2019PhRvD.100d3027R}
{Rodriguez}, C.~L., {Zevin}, M., {Amaro-Seoane}, P., {et~al.} 2019, \prd, 100,
  043027, \dodoi{10.1103/PhysRevD.100.043027}

\bibitem[{{Rodriguez} {et~al.}(2016{\natexlab{c}}){Rodriguez}, {Zevin},
  {Pankow}, {Kalogera}, \& {Rasio}}]{2016ApJ...832L...2R}
{Rodriguez}, C.~L., {Zevin}, M., {Pankow}, C., {Kalogera}, V., \& {Rasio},
  F.~A. 2016{\natexlab{c}}, \apjl, 832, L2, \dodoi{10.3847/2041-8205/832/1/L2}

\bibitem[{{Safarzadeh} {et~al.}(2020){Safarzadeh}, {Hamers}, {Loeb}, \&
  {Berger}}]{2020ApJ...888L...3S}
{Safarzadeh}, M., {Hamers}, A.~S., {Loeb}, A., \& {Berger}, E. 2020, \apjl,
  888, L3, \dodoi{10.3847/2041-8213/ab5dc8}

\bibitem[{{Samsing}(2018)}]{2018PhRvD..97j3014S}
{Samsing}, J. 2018, \prd, 97, 103014, \dodoi{10.1103/PhysRevD.97.103014}

\bibitem[{{Samsing} {et~al.}(2018{\natexlab{a}}){Samsing}, {Askar}, \&
  {Giersz}}]{2018ApJ...855..124S}
{Samsing}, J., {Askar}, A., \& {Giersz}, M. 2018{\natexlab{a}}, \apj, 855, 124,
  \dodoi{10.3847/1538-4357/aaab52}

\bibitem[{{Samsing} \& {D'Orazio}(2018)}]{2018MNRAS.tmp.2223S}
{Samsing}, J., \& {D'Orazio}, D.~J. 2018, \mnras, \dodoi{10.1093/mnras/sty2334}

\bibitem[{{Samsing} {et~al.}(2019{\natexlab{a}}){Samsing}, {D'Orazio},
  {Kremer}, {Rodriguez}, \& {Askar}}]{2019arXiv190711231S}
{Samsing}, J., {D'Orazio}, D.~J., {Kremer}, K., {Rodriguez}, C.~L., \& {Askar},
  A. 2019{\natexlab{a}}, arXiv e-prints, arXiv:1907.11231.
\newblock \doarXiv{1907.11231}

\bibitem[{{Samsing} {et~al.}(2019{\natexlab{b}}){Samsing}, {Hamers}, \&
  {Tyles}}]{2019PhRvD.100d3010S}
{Samsing}, J., {Hamers}, A.~S., \& {Tyles}, J.~G. 2019{\natexlab{b}}, \prd,
  100, 043010, \dodoi{10.1103/PhysRevD.100.043010}

\bibitem[{{Samsing} \& {Ilan}(2018)}]{2018MNRAS.476.1548S}
{Samsing}, J., \& {Ilan}, T. 2018, \mnras, 476, 1548,
  \dodoi{10.1093/mnras/sty197}

\bibitem[{{Samsing} \& {Ilan}(2019)}]{2019MNRAS.482...30S}
---. 2019, \mnras, 482, 30, \dodoi{10.1093/mnras/sty2249}

\bibitem[{{Samsing} {et~al.}(2014){Samsing}, {MacLeod}, \&
  {Ramirez-Ruiz}}]{2014ApJ...784...71S}
{Samsing}, J., {MacLeod}, M., \& {Ramirez-Ruiz}, E. 2014, \apj, 784, 71,
  \dodoi{10.1088/0004-637X/784/1/71}

\bibitem[{{Samsing} {et~al.}(2018{\natexlab{b}}){Samsing}, {MacLeod}, \&
  {Ramirez-Ruiz}}]{2018ApJ...853..140S}
---. 2018{\natexlab{b}}, \apj, 853, 140, \dodoi{10.3847/1538-4357/aaa715}

\bibitem[{{Samsing} \& {Ramirez-Ruiz}(2017)}]{2017ApJ...840L..14S}
{Samsing}, J., \& {Ramirez-Ruiz}, E. 2017, \apjl, 840, L14,
  \dodoi{10.3847/2041-8213/aa6f0b}

\bibitem[{{Samsing} {et~al.}(2019{\natexlab{c}}){Samsing}, {Venumadhav}, {Dai},
  {Martinez}, {Batta}, {Lopez}, {Ramirez-Ruiz}, \&
  {Kremer}}]{2019PhRvD.100d3009S}
{Samsing}, J., {Venumadhav}, T., {Dai}, L., {et~al.} 2019{\natexlab{c}}, \prd,
  100, 043009, \dodoi{10.1103/PhysRevD.100.043009}

\bibitem[{{Sasaki} {et~al.}(2016){Sasaki}, {Suyama}, {Tanaka}, \&
  {Yokoyama}}]{2016PhRvL.117f1101S}
{Sasaki}, M., {Suyama}, T., {Tanaka}, T., \& {Yokoyama}, S. 2016, Physical
  Review Letters, 117, 061101, \dodoi{10.1103/PhysRevLett.117.061101}

\bibitem[{{Schr{\o}der} {et~al.}(2018){Schr{\o}der}, {Batta}, \&
  {Ramirez-Ruiz}}]{2018ApJ...862L...3S}
{Schr{\o}der}, S.~L., {Batta}, A., \& {Ramirez-Ruiz}, E. 2018, \apjl, 862, L3,
  \dodoi{10.3847/2041-8213/aacf8d}

\bibitem[{Sigurdsson \& Phinney(1993)}]{Sigurdsson:1993jz}
Sigurdsson, S., \& Phinney, E.~S. 1993, \apj, 415, 631

\bibitem[{{Silsbee} \& {Tremaine}(2017)}]{2017ApJ...836...39S}
{Silsbee}, K., \& {Tremaine}, S. 2017, \apj, 836, 39,
  \dodoi{10.3847/1538-4357/aa5729}

\bibitem[{{Stephan} {et~al.}(2016){Stephan}, {Naoz}, {Ghez}, {Witzel},
  {Sitarski}, {Do}, \& {Kocsis}}]{2016MNRAS.460.3494S}
{Stephan}, A.~P., {Naoz}, S., {Ghez}, A.~M., {et~al.} 2016, \mnras, 460, 3494,
  \dodoi{10.1093/mnras/stw1220}

\bibitem[{{Stone} \& {Leigh}(2019)}]{2019Natur.576..406S}
{Stone}, N.~C., \& {Leigh}, N. W.~C. 2019, \nat, 576, 406,
  \dodoi{10.1038/s41586-019-1833-8}

\bibitem[{{Stone} {et~al.}(2017){Stone}, {Metzger}, \&
  {Haiman}}]{2017MNRAS.464..946S}
{Stone}, N.~C., {Metzger}, B.~D., \& {Haiman}, Z. 2017, \mnras, 464, 946,
  \dodoi{10.1093/mnras/stw2260}

\bibitem[{{Tagawa} {et~al.}(2019){Tagawa}, {Haiman}, \&
  {Kocsis}}]{2019arXiv191208218T}
{Tagawa}, H., {Haiman}, Z., \& {Kocsis}, B. 2019, arXiv e-prints,
  arXiv:1912.08218.
\newblock \doarXiv{1912.08218}

\bibitem[{{Tanikawa}(2013)}]{2013MNRAS.435.1358T}
{Tanikawa}, A. 2013, \mnras, 435, 1358, \dodoi{10.1093/mnras/stt1380}

\bibitem[{{Toonen} {et~al.}(2016){Toonen}, {Hamers}, \& {Portegies
  Zwart}}]{2016ComAC...3....6T}
{Toonen}, S., {Hamers}, A., \& {Portegies Zwart}, S. 2016, Computational
  Astrophysics and Cosmology, 3, 6, \dodoi{10.1186/s40668-016-0019-0}

\bibitem[{{VanLandingham} {et~al.}(2016){VanLandingham}, {Miller}, {Hamilton},
  \& {Richardson}}]{2016ApJ...828...77V}
{VanLandingham}, J.~H., {Miller}, M.~C., {Hamilton}, D.~P., \& {Richardson},
  D.~C. 2016, \apj, 828, 77, \dodoi{10.3847/0004-637X/828/2/77}

\bibitem[{{Venumadhav} {et~al.}(2019){Venumadhav}, {Zackay}, {Roulet}, {Dai},
  \& {Zaldarriaga}}]{2019arXiv190407214V}
{Venumadhav}, T., {Zackay}, B., {Roulet}, J., {Dai}, L., \& {Zaldarriaga}, M.
  2019, arXiv e-prints, arXiv:1904.07214.
\newblock \doarXiv{1904.07214}

\bibitem[{{Woosley}(2016)}]{Woosley:2016}
{Woosley}, S.~E. 2016, \apjl, 824, L10, \dodoi{10.3847/2041-8205/824/1/L10}

\bibitem[{{Woosley}(2017)}]{Woosley2017ApJ}
---. 2017, \apj, 836, 244, \dodoi{10.3847/1538-4357/836/2/244}

\bibitem[{{Yang} {et~al.}(2019){Yang}, {Bartos}, {Gayathri}, {Ford}, {Haiman},
  {Klimenko}, {Kocsis}, {M{\'a}rka}, {M{\'a}rka}, {McKernan}, \&
  {O'Shaughnessy}}]{2019PhRvL.123r1101Y}
{Yang}, Y., {Bartos}, I., {Gayathri}, V., {et~al.} 2019, \prl, 123, 181101,
  \dodoi{10.1103/PhysRevLett.123.181101}

\bibitem[{{Ye} {et~al.}(2020){Ye}, {Fong}, {Kremer}, {Rodriguez}, {Chatterjee},
  {Fragione}, \& {Rasio}}]{2020ApJ...888L..10Y}
{Ye}, C.~S., {Fong}, W.-f., {Kremer}, K., {et~al.} 2020, \apjl, 888, L10,
  \dodoi{10.3847/2041-8213/ab5dc5}

\bibitem[{{Zackay} {et~al.}(2019){Zackay}, {Venumadhav}, {Dai}, {Roulet}, \&
  {Zaldarriaga}}]{2019arXiv190210331Z}
{Zackay}, B., {Venumadhav}, T., {Dai}, L., {Roulet}, J., \& {Zaldarriaga}, M.
  2019, arXiv e-prints, arXiv:1902.10331.
\newblock \doarXiv{1902.10331}

\bibitem[{{Zaldarriaga} {et~al.}(2018){Zaldarriaga}, {Kushnir}, \&
  {Kollmeier}}]{Zaldarriaga2018MNRAS}
{Zaldarriaga}, M., {Kushnir}, D., \& {Kollmeier}, J.~A. 2018, \mnras, 473,
  4174, \dodoi{10.1093/mnras/stx2577}

\bibitem[{{Zevin} {et~al.}(2017){Zevin}, {Pankow}, {Rodriguez}, {Sampson},
  {Chase}, {Kalogera}, \& {Rasio}}]{2017ApJ...846...82Z}
{Zevin}, M., {Pankow}, C., {Rodriguez}, C.~L., {et~al.} 2017, \apj, 846, 82,
  \dodoi{10.3847/1538-4357/aa8408}

\bibitem[{{Zevin} {et~al.}(2019){Zevin}, {Samsing}, {Rodriguez}, {Haster}, \&
  {Ramirez-Ruiz}}]{2019ApJ...871...91Z}
{Zevin}, M., {Samsing}, J., {Rodriguez}, C., {Haster}, C.-J., \&
  {Ramirez-Ruiz}, E. 2019, \apj, 871, 91, \dodoi{10.3847/1538-4357/aaf6ec}

\end{thebibliography}

\end{document}